\documentclass[aps, prd,twocolumn,superscriptaddress,nofootinbib,preprintnumbers]{revtex4-1}

\def\araa{ARA\&A}%
\def\apj{ApJ}%
\def\apjs{ApJS}%
\def\aap{A\&A}%
\def\mnras{MNRAS}%
\def\prd{Phys.~Rev.~D}%
\def\pasp{PASP}%
\def\jcap{JCAP}%
\def\physrep{Phys.~Rep.}%
\newcommand{\om}{\Omega_\mr m}
\newcommand{\omb}{\Omega_\mr b}

\newcommand{\w}{w_0}

\def\galk{w^{\delta_{\rm g} \kappa_{\rm CMB}}(\theta) }
\def\galik{w^{\delta_{\rm g}^i \kappa_{\rm CMB}}(\theta) }
\def\sheark{w^{\gamma \kappa_{\rm CMB}}(\theta)}
\def\sheartk{w^{\gamma_{\rm t} \kappa_{\rm CMB}}(\theta)}
\def\sheartik{w^{\gamma^i_{\rm t} \kappa_{\rm CMB}}(\theta)}
\def\kk{w^{\kappa_{\rm CMB}\kappa_{\rm CMB}}(\theta)}
\def\galkobs{w^{\delta_{\rm g} \kappa_{\rm obs}}(\theta)}
\def\galksim{w^{\delta_{\rm g} \kappa_{\rm sim}}(\theta)}
\def\galksys{w^{\delta_{\rm g} \kappa_{\rm sys}}(\theta)}

\def\sheartkobs{w^{\gamma_{\rm t} \kappa_{\rm obs}}(\theta)}
\def\sheartksim{w^{\gamma_{\rm t} \kappa_{\rm sim}}(\theta)}

\def\sheartksys{w^{\gamma_{\rm t} \kappa_{\rm sys}}(\theta)}

\def\galgal{w^{\delta_{\rm g} \delta_{\rm g}}(\theta)}
\def\galshear{w^{\delta_{\rm g} \gamma}(\theta)}
\def\galsheart{w^{\delta_{\rm g} \gamma_{\rm t}}(\theta)}
\def\shearshear{w^{\gamma \gamma}(\theta)}
\def\wtheta{w^{\delta_{\rm g} \delta_{\rm g}}(\theta)}

\def\galkell{C^{\delta_{\rm g} \kappa_{\rm CMB}} (\ell)}
\def\galikell{C^{\delta_{\rm g}^i \kappa_{\rm CMB}} (\ell)}
\def\galktszell{C^{\delta_{\rm g} \kappa_{\rm tSZ}} (\ell)}
\def\kappaskell{C^{\kappa_{\rm s} \kappa_{\rm CMB}}(\ell)}
\def\kappasikell{C^{\kappa_{\rm s} \kappa_{\rm CMB}}(\ell)}
\def\kappasktszell{C^{\kappa_{\rm s} \kappa_{\rm tSZ}}(\ell)}

\def\3x2pt{3$\times$2pt}
\def\5x2pt{5$\times$2pt}
\def\6x2pt{6$\times$2pt}

\newcommand{\kcmb}{\kappa_{\rm CMB}}
\newcommand{\ktsz}{\kappa_{\rm tSZ}}

\newcommand{\delg}{\delta_{\rm g}}

\usepackage{amsmath,amssymb,latexsym,times}
\usepackage{numdef}
\usepackage{slashed}
\usepackage{graphicx}
\usepackage{grffile}
\usepackage[normalem]{ulem}
\usepackage{dcolumn}
\usepackage{xcolor}
\usepackage{booktabs}
\usepackage{amssymb}
\usepackage{slashed}
\usepackage{todonotes}
\definecolor{aiiro}{HTML}{952525}
\usepackage{hyperref}
 \hypersetup{
     colorlinks=true,
     linkcolor=blue,
     citecolor = blue,      
     urlcolor=blue,
     }

\usepackage{dcolumn}
\usepackage{xcolor}
\usepackage{booktabs}
\usepackage{amsmath}
\usepackage{amssymb}
\usepackage{slashed}
\usepackage{todonotes}
\usepackage{multirow}
\usepackage{hyperref}
\usepackage[T1]{fontenc}
\usepackage{ae,aecompl}

\usepackage{lineno}

\usepackage{eso-pic}

\usepackage{txfonts}

\newcommand{\nside}{\ifmmode N_{\mathrm{side}}\else $N_{\mathrm{side}}$\fi}
\newcommand{\npix}{\ifmmode n_{\mathrm{pix}}\else $n_{\mathrm{pix}}$\fi}
\newcommand{\Npix}{\ifmmode N_{\mathrm{pix}}\else $n_{\mathrm{pix}}$\fi}
\newcommand{\lmin}{\ifmmode \ell_{\mathrm{min}}\else $\ell_{\mathrm{min}}$\fi}
\newcommand{\lmax}{\ifmmode \ell_{\mathrm{max}}\else $\ell_{\mathrm{max}}$\fi}

\newcommand{\Planck}{{\slshape Planck~}}
\newcommand{\Planckc}{{\slshape Planck}}

\newcommand{\mr}[1]{\mathrm{#1}}

\defcitealias{Krause:2017}{K17}
\defcitealias{Melchior:2017}{M17}
\defcitealias{Omori:2017}{O17}

\begin{document}

\title[\5x2pt Methodology]
{Dark Energy Survey Year 1 Results: Methodology and Projections for Joint Analysis of Galaxy Clustering, Galaxy Lensing, and CMB Lensing Two-point Functions}

\author{E.~J.~Baxter}
\affiliation{Department of Physics and Astronomy, University of Pennsylvania, Philadelphia, PA 19104, USA}
\author{Y.~Omori}
\affiliation{Kavli Institute for Particle Astrophysics and Cosmology, Stanford University, 452 Lomita Mall, Stanford, CA 94305}
\affiliation{Dept. of Physics, Stanford University, 382 Via Pueblo Mall, Stanford, CA 94305}
\affiliation{Department of Physics and McGill Space Institute, McGill University, Montreal, Quebec H3A 2T8, Canada}
\author{C.~Chang}
\affiliation{Kavli Institute for Cosmological Physics, University of Chicago, Chicago, IL 60637, USA}
\author{T.~Giannantonio}
\affiliation{Institute of Astronomy, University of Cambridge, Madingley Road, Cambridge CB3 0HA, UK}
\affiliation{Kavli Institute for Cosmology, University of Cambridge, Madingley Road, Cambridge CB3 0HA, UK}
\affiliation{Universit\"ats-Sternwarte, Fakult\"at f\"ur Physik, Ludwig-Maximilians Universit\"at M\"unchen, Scheinerstr. 1, 81679 M\"unchen, Germany}
\author{D.~Kirk}
\affiliation{Department of Physics \& Astronomy, University College London, Gower Street, London, WC1E 6BT, UK}
\author{E.~Krause}
\affiliation{Department of Astronomy/Steward Observatory, 933 North Cherry Avenue, Tucson, AZ 85721-0065, USA}
\affiliation{Jet Propulsion Laboratory, California Institute of Technology, 4800 Oak Grove Dr., Pasadena, CA 91109, USA}
\author{J.~Blazek}
\affiliation{Center for Cosmology and Astro-Particle Physics, The Ohio State University, Columbus, OH 43210, USA}
\affiliation{Institute of Physics, Laboratory of Astrophysics, \'Ecole Polytechnique F\'ed\'erale de Lausanne (EPFL), Observatoire de Sauverny, 1290 Versoix, Switzerland}
\author{L.~Bleem}
\affiliation{Department of Physics, University of Chicago, 5640 South Ellis Avenue, Chicago, IL 60637, USA}
\affiliation{Argonne National Laboratory, 9700 South Cass Avenue, Lemont, IL 60439, USA}
\author{A.~Choi}
\affiliation{Scottish Universities Physics Alliance, Institute for Astronomy, University of Edinburgh, Royal Observatory, Blackford Hill, Edinburgh, EH9 3HJ, UK}
\affiliation{Center for Cosmology and AstroParticle Physics, The Ohio State University, 191 West Woodruff Avenue, Columbus, OH 43210, USA}
\author{T.~M.~Crawford}
\affiliation{Kavli Institute for Cosmological Physics, University of Chicago, Chicago, IL 60637, USA}
\affiliation{Department of Astronomy and Astrophysics, University of Chicago, Chicago, IL 60637, USA}
\author{S.~Dodelson}
\affiliation{Observatories of the Carnegie Institution of Washington, 813 Santa Barbara St., Pasadena, CA 91101, USA}
\author{D.~Gruen}
\affiliation{Kavli Institute for Particle Astrophysics \& Cosmology, P. O. Box 2450, Stanford University, Stanford, CA 94305, USA}
\affiliation{SLAC National Accelerator Laboratory, Menlo Park, CA 94025, USA}
\author{G.~P.~Holder}
\affiliation{Canadian Institute for Advanced Research, CIFAR Program on Gravity and the Extreme Universe, Toronto, ON, M5G 1Z8, Canada}
\affiliation{Astronomy Department, University of Illinois at Urbana Champaign, 1002 W. Green Street, Urbana, IL 61801, USA}
\affiliation{Department of Physics, University of Illinois Urbana Champaign, 1110 W. Green Street, Urbana, IL 61801, USA}
\author{B.~Jain}
\affiliation{Department of Physics and Astronomy, University of Pennsylvania, Philadelphia, PA 19104, USA}
\author{M.~Jarvis}
\affiliation{Department of Physics and Astronomy, University of Pennsylvania, Philadelphia, PA 19104, USA}
\author{N.~MacCrann}
\affiliation{Center for Cosmology and Astro-Particle Physics, The Ohio State University, Columbus, OH 43210, USA}
\affiliation{Department of Physics, The Ohio State University, Columbus, OH 43210, USA}
\author{A.~Nicola}
\affiliation{Department of Physics, ETH Zurich, Wolfgang-Pauli-Strasse 16, CH-8093 Zurich, Switzerland}
\author{S.~Pandey}
\affiliation{Department of Physics and Astronomy, University of Pennsylvania, Philadelphia, PA 19104, USA}
\author{J.~Prat}
\affiliation{Institut de F\'{\i}sica d'Altes Energies (IFAE), The Barcelona Institute of Science and Technology, Campus UAB, 08193 Bellaterra (Barcelona) Spain}
\author{S.~Samuroff}
\affiliation{Jodrell Bank Center for Astrophysics, School of Physics and Astronomy, University of Manchester, Oxford Road, Manchester, M13 9PL, UK}
\author{C.~S{\'a}nchez}
\affiliation{Institut de F\'{\i}sica d'Altes Energies (IFAE), The Barcelona Institute of Science and Technology, Campus UAB, 08193 Bellaterra (Barcelona) Spain}
\author{E.~Sheldon}
\affiliation{Brookhaven National Laboratory, Bldg 510, Upton, NY 11973, USA}
\author{M.~A.~Troxel}
\affiliation{Center for Cosmology and Astro-Particle Physics, The Ohio State University, Columbus, OH 43210, USA}
\affiliation{Department of Physics, The Ohio State University, Columbus, OH 43210, USA}
\author{J.~Zuntz}
\affiliation{Institute for Astronomy, University of Edinburgh, Edinburgh EH9 3HJ, UK}
\author{T.~M.~C.~Abbott}
\affiliation{Cerro Tololo Inter-American Observatory, National Optical Astronomy Observatory, Casilla 603, La Serena, Chile}
\author{F.~B.~Abdalla}
\affiliation{Department of Physics \& Astronomy, University College London, Gower Street, London, WC1E 6BT, UK}
\affiliation{Department of Physics and Electronics, Rhodes University, PO Box 94, Grahamstown, 6140, South Africa}
\author{J.~Annis}
\affiliation{Fermi National Accelerator Laboratory, P. O. Box 500, Batavia, IL 60510, USA}
\author{S.~Avila}
\affiliation{Institute of Cosmology \& Gravitation, University of Portsmouth, Portsmouth, PO1 3FX, UK}
\author{K.~Bechtol}
\affiliation{LSST, 933 North Cherry Avenue, Tucson, AZ 85721, USA}
\author{E.~Bertin}
\affiliation{CNRS, UMR 7095, Institut d'Astrophysique de Paris, F-75014, Paris, France}
\affiliation{Sorbonne Universit\'es, UPMC Univ Paris 06, UMR 7095, Institut d'Astrophysique de Paris, F-75014, Paris, France}
\author{D.~Brooks}
\affiliation{Department of Physics \& Astronomy, University College London, Gower Street, London, WC1E 6BT, UK}
\author{E.~Buckley-Geer}
\affiliation{Fermi National Accelerator Laboratory, P. O. Box 500, Batavia, IL 60510, USA}
\author{D.~L.~Burke}
\affiliation{Kavli Institute for Particle Astrophysics \& Cosmology, P. O. Box 2450, Stanford University, Stanford, CA 94305, USA}
\affiliation{SLAC National Accelerator Laboratory, Menlo Park, CA 94025, USA}
\author{A.~Carnero~Rosell}
\affiliation{Laborat\'orio Interinstitucional de e-Astronomia - LIneA, Rua Gal. Jos\'e Cristino 77, Rio de Janeiro, RJ - 20921-400, Brazil}
\affiliation{Observat\'orio Nacional, Rua Gal. Jos\'e Cristino 77, Rio de Janeiro, RJ - 20921-400, Brazil}
\author{M.~Carrasco~Kind}
\affiliation{Department of Astronomy, University of Illinois at Urbana-Champaign, 1002 W. Green Street, Urbana, IL 61801, USA}
\affiliation{National Center for Supercomputing Applications, 1205 West Clark St., Urbana, IL 61801, USA}
\author{J.~Carretero}
\affiliation{Institut de F\'{\i}sica d'Altes Energies (IFAE), The Barcelona Institute of Science and Technology, Campus UAB, 08193 Bellaterra (Barcelona) Spain}
\author{F.~J.~Castander}
\affiliation{Institut d'Estudis Espacials de Catalunya (IEEC), 08193 Barcelona, Spain}
\affiliation{Institute of Space Sciences (ICE, CSIC),  Campus UAB, Carrer de Can Magrans, s/n,  08193 Barcelona, Spain}
\author{R.~Cawthon}
\affiliation{Kavli Institute for Cosmological Physics, University of Chicago, Chicago, IL 60637, USA}
\author{C.~E.~Cunha}
\affiliation{Kavli Institute for Particle Astrophysics \& Cosmology, P. O. Box 2450, Stanford University, Stanford, CA 94305, USA}
\author{C.~B.~D'Andrea}
\affiliation{Department of Physics and Astronomy, University of Pennsylvania, Philadelphia, PA 19104, USA}
\author{L.~N.~da Costa}
\affiliation{Laborat\'orio Interinstitucional de e-Astronomia - LIneA, Rua Gal. Jos\'e Cristino 77, Rio de Janeiro, RJ - 20921-400, Brazil}
\affiliation{Observat\'orio Nacional, Rua Gal. Jos\'e Cristino 77, Rio de Janeiro, RJ - 20921-400, Brazil}
\author{C.~Davis}
\affiliation{Kavli Institute for Particle Astrophysics \& Cosmology, P. O. Box 2450, Stanford University, Stanford, CA 94305, USA}
\author{J.~De~Vicente}
\affiliation{Centro de Investigaciones Energ\'eticas, Medioambientales y Tecnol\'ogicas (CIEMAT), Madrid, Spain}
\author{D.~L.~DePoy}
\affiliation{George P. and Cynthia Woods Mitchell Institute for Fundamental Physics and Astronomy, and Department of Physics and Astronomy, Texas A\&M University, College Station, TX 77843,  USA}
\author{H.~T.~Diehl}
\affiliation{Fermi National Accelerator Laboratory, P. O. Box 500, Batavia, IL 60510, USA}
\author{P.~Doel}
\affiliation{Department of Physics \& Astronomy, University College London, Gower Street, London, WC1E 6BT, UK}
\author{T.~F.~Eifler}
\affiliation{Department of Astronomy/Steward Observatory, 933 North Cherry Avenue, Tucson, AZ 85721-0065, USA}
\affiliation{Jet Propulsion Laboratory, California Institute of Technology, 4800 Oak Grove Dr., Pasadena, CA 91109, USA}
\author{J.~Estrada}
\affiliation{Fermi National Accelerator Laboratory, P. O. Box 500, Batavia, IL 60510, USA}
\author{A.~E.~Evrard}
\affiliation{Department of Astronomy, University of Michigan, Ann Arbor, MI 48109, USA}
\affiliation{Department of Physics, University of Michigan, Ann Arbor, MI 48109, USA}
\author{B.~Flaugher}
\affiliation{Fermi National Accelerator Laboratory, P. O. Box 500, Batavia, IL 60510, USA}
\author{P.~Fosalba}
\affiliation{Institut d'Estudis Espacials de Catalunya (IEEC), 08193 Barcelona, Spain}
\affiliation{Institute of Space Sciences (ICE, CSIC),  Campus UAB, Carrer de Can Magrans, s/n,  08193 Barcelona, Spain}
\author{J.~Frieman}
\affiliation{Fermi National Accelerator Laboratory, P. O. Box 500, Batavia, IL 60510, USA}
\affiliation{Kavli Institute for Cosmological Physics, University of Chicago, Chicago, IL 60637, USA}
\author{J.~Garc\'ia-Bellido}
\affiliation{Instituto de Fisica Teorica UAM/CSIC, Universidad Autonoma de Madrid, 28049 Madrid, Spain}
\author{E.~Gaztanaga}
\affiliation{Institut d'Estudis Espacials de Catalunya (IEEC), 08193 Barcelona, Spain}
\affiliation{Institute of Space Sciences (ICE, CSIC),  Campus UAB, Carrer de Can Magrans, s/n,  08193 Barcelona, Spain}
\author{D.~W.~Gerdes}
\affiliation{Department of Astronomy, University of Michigan, Ann Arbor, MI 48109, USA}
\affiliation{Department of Physics, University of Michigan, Ann Arbor, MI 48109, USA}
\author{R.~A.~Gruendl}
\affiliation{Department of Astronomy, University of Illinois at Urbana-Champaign, 1002 W. Green Street, Urbana, IL 61801, USA}
\affiliation{National Center for Supercomputing Applications, 1205 West Clark St., Urbana, IL 61801, USA}
\author{J.~Gschwend}
\affiliation{Laborat\'orio Interinstitucional de e-Astronomia - LIneA, Rua Gal. Jos\'e Cristino 77, Rio de Janeiro, RJ - 20921-400, Brazil}
\affiliation{Observat\'orio Nacional, Rua Gal. Jos\'e Cristino 77, Rio de Janeiro, RJ - 20921-400, Brazil}
\author{G.~Gutierrez}
\affiliation{Fermi National Accelerator Laboratory, P. O. Box 500, Batavia, IL 60510, USA}
\author{W.~G.~Hartley}
\affiliation{Department of Physics \& Astronomy, University College London, Gower Street, London, WC1E 6BT, UK}
\affiliation{Department of Physics, ETH Zurich, Wolfgang-Pauli-Strasse 16, CH-8093 Zurich, Switzerland}
\author{D.~Hollowood}
\affiliation{Santa Cruz Institute for Particle Physics, Santa Cruz, CA 95064, USA}
\author{B.~Hoyle}
\affiliation{Max Planck Institute for Extraterrestrial Physics, Giessenbachstrasse, 85748 Garching, Germany}
\affiliation{Universit\"ats-Sternwarte, Fakult\"at f\"ur Physik, Ludwig-Maximilians Universit\"at M\"unchen, Scheinerstr. 1, 81679 M\"unchen, Germany}
\author{D.~J.~James}
\affiliation{Harvard-Smithsonian Center for Astrophysics, Cambridge, MA 02138, USA}
\author{S.~Kent}
\affiliation{Fermi National Accelerator Laboratory, P. O. Box 500, Batavia, IL 60510, USA}
\affiliation{Kavli Institute for Cosmological Physics, University of Chicago, Chicago, IL 60637, USA}
\author{K.~Kuehn}
\affiliation{Australian Astronomical Observatory, North Ryde, NSW 2113, Australia}
\author{N.~Kuropatkin}
\affiliation{Fermi National Accelerator Laboratory, P. O. Box 500, Batavia, IL 60510, USA}
\author{O.~Lahav}
\affiliation{Department of Physics \& Astronomy, University College London, Gower Street, London, WC1E 6BT, UK}
\author{M.~Lima}
\affiliation{Departamento de F\'isica Matem\'atica, Instituto de F\'isica, Universidade de S\~ao Paulo, CP 66318, S\~ao Paulo, SP, 05314-970, Brazil}
\affiliation{Laborat\'orio Interinstitucional de e-Astronomia - LIneA, Rua Gal. Jos\'e Cristino 77, Rio de Janeiro, RJ - 20921-400, Brazil}
\author{M.~A.~G.~Maia}
\affiliation{Laborat\'orio Interinstitucional de e-Astronomia - LIneA, Rua Gal. Jos\'e Cristino 77, Rio de Janeiro, RJ - 20921-400, Brazil}
\affiliation{Observat\'orio Nacional, Rua Gal. Jos\'e Cristino 77, Rio de Janeiro, RJ - 20921-400, Brazil}
\author{M.~March}
\affiliation{Department of Physics and Astronomy, University of Pennsylvania, Philadelphia, PA 19104, USA}
\author{J.~L.~Marshall}
\affiliation{George P. and Cynthia Woods Mitchell Institute for Fundamental Physics and Astronomy, and Department of Physics and Astronomy, Texas A\&M University, College Station, TX 77843,  USA}
\author{P.~Melchior}
\affiliation{Department of Astrophysical Sciences, Princeton University, Peyton Hall, Princeton, NJ 08544, USA}
\author{F.~Menanteau}
\affiliation{Department of Astronomy, University of Illinois at Urbana-Champaign, 1002 W. Green Street, Urbana, IL 61801, USA}
\affiliation{National Center for Supercomputing Applications, 1205 West Clark St., Urbana, IL 61801, USA}
\author{R.~Miquel}
\affiliation{Instituci\'o Catalana de Recerca i Estudis Avan\c{c}ats, E-08010 Barcelona, Spain}
\affiliation{Institut de F\'{\i}sica d'Altes Energies (IFAE), The Barcelona Institute of Science and Technology, Campus UAB, 08193 Bellaterra (Barcelona) Spain}
\author{A.~A.~Plazas}
\affiliation{Jet Propulsion Laboratory, California Institute of Technology, 4800 Oak Grove Dr., Pasadena, CA 91109, USA}
\author{A.~Roodman}
\affiliation{Kavli Institute for Particle Astrophysics \& Cosmology, P. O. Box 2450, Stanford University, Stanford, CA 94305, USA}
\affiliation{SLAC National Accelerator Laboratory, Menlo Park, CA 94025, USA}
\author{E.~S.~Rykoff}
\affiliation{Kavli Institute for Particle Astrophysics \& Cosmology, P. O. Box 2450, Stanford University, Stanford, CA 94305, USA}
\affiliation{SLAC National Accelerator Laboratory, Menlo Park, CA 94025, USA}
\author{E.~Sanchez}
\affiliation{Centro de Investigaciones Energ\'eticas, Medioambientales y Tecnol\'ogicas (CIEMAT), Madrid, Spain}
\author{R.~Schindler}
\affiliation{SLAC National Accelerator Laboratory, Menlo Park, CA 94025, USA}
\author{M.~Schubnell}
\affiliation{Department of Physics, University of Michigan, Ann Arbor, MI 48109, USA}
\author{I.~Sevilla-Noarbe}
\affiliation{Centro de Investigaciones Energ\'eticas, Medioambientales y Tecnol\'ogicas (CIEMAT), Madrid, Spain}
\author{M.~Smith}
\affiliation{School of Physics and Astronomy, University of Southampton,  Southampton, SO17 1BJ, UK}
\author{R.~C.~Smith}
\affiliation{Cerro Tololo Inter-American Observatory, National Optical Astronomy Observatory, Casilla 603, La Serena, Chile}
\author{M.~Soares-Santos}
\affiliation{Center for Particle Astrophysics, Fermi National Accelerator Laboratory, Batavia, IL 60510, USA}
\author{F.~Sobreira}
\affiliation{Instituto de F\'isica Gleb Wataghin, Universidade Estadual de Campinas, 13083-859, Campinas, SP, Brazil}
\affiliation{Laborat\'orio Interinstitucional de e-Astronomia - LIneA, Rua Gal. Jos\'e Cristino 77, Rio de Janeiro, RJ - 20921-400, Brazil}
\author{E.~Suchyta}
\affiliation{Computer Science and Mathematics Division, Oak Ridge National Laboratory, Oak Ridge, TN 37831}
\author{M.~E.~C.~Swanson}
\affiliation{National Center for Supercomputing Applications, 1205 West Clark St., Urbana, IL 61801, USA}
\author{G.~Tarle}
\affiliation{Department of Physics, University of Michigan, Ann Arbor, MI 48109, USA}
\author{A.~R.~Walker}
\affiliation{Cerro Tololo Inter-American Observatory, National Optical Astronomy Observatory, Casilla 603, La Serena, Chile}
\author{J.~Weller}
\affiliation{Excellence Cluster Universe, Boltzmannstr.\ 2, 85748 Garching, Germany}
\affiliation{Max Planck Institute for Extraterrestrial Physics, Giessenbachstrasse, 85748 Garching, Germany}
\affiliation{Universit\"ats-Sternwarte, Fakult\"at f\"ur Physik, Ludwig-Maximilians Universit\"at M\"unchen, Scheinerstr. 1, 81679 M\"unchen, Germany}

\collaboration{DES and SPT Collaborations}

\date{Last updated \today}


\begin{abstract}
Optical imaging surveys measure both the galaxy density and the gravitational lensing-induced shear fields across the sky.  Recently, the Dark Energy Survey (DES) collaboration used a joint fit to two-point correlations between these observables to place tight constraints on cosmology (DES Collaboration {\it et al.} 2018).  In this work, we develop the methodology to extend the DES year one joint probes analysis to include cross-correlations of the optical survey observables with gravitational lensing of the cosmic microwave background (CMB) as measured by the South Pole Telescope (SPT) and \Planckc.  Using simulated analyses, we show how the resulting set of five two-point functions increases the robustness of the cosmological constraints to systematic errors in galaxy lensing shear calibration.  Additionally, we show that contamination of the SPT+\Planck CMB lensing map by the thermal Sunyaev-Zel'dovich effect is a potentially large source of systematic error for two-point function analyses, but show that it can be reduced to acceptable levels in our analysis by masking clusters of galaxies and imposing angular scale cuts on the two-point functions.  The methodology developed here will be applied to the analysis of data from the DES, the SPT, and \Planck in a companion work.
\end{abstract}

\preprint{DES-2018-0294}
\preprint{FERMILAB-PUB-18-045-A-AE}

\maketitle

\section{Introduction}

Modern optical imaging surveys measure the positions and gravitational lensing-induced shears of millions of galaxies. From these measurements, one can compute two fields on the sky: the spin-0 galaxy overdensity field, $\delta_{\rm g}$, and the spin-2 weak lensing shear field, $\gamma$. Two-point cross-correlations between these fields are powerful cosmological probes, as they are sensitive to both the geometry of the Universe and the growth of structure. Joint fits to multiple two-point correlations --- such as $\delg\delg$ and $\delg\gamma$ --- offer the possibility of breaking degeneracies between cosmological and nuisance parameters, as well as significantly improving cosmological constraints \citep[e.g.][]{Hu:2004}.\footnote{We will use the notation $w^{XY}(\theta)$ to represent the configuration space, two-point correlation function between fields $X$ and $Y$.  We will use the notation $C^{XY}(\ell)$ to represent the harmonic space cross-power spectrum between two fields.} Such joint fits have recently been demonstrated in several works \citep{Nicola2016,Kwan:2017, vanUitert:2017,DESy1:2017,Joudaki:2018}. We refer to the set of three two-point functions that can be formed from $\gamma$ and $\delta_{\rm g}$ --- namely $\delg\delg$, $\delg\gamma$, and $\gamma\gamma$ --- as \3x2pt. The \3x2pt analysis of the \citet{DESy1:2017} presented the tightest cosmological constraints to date on $\om$ and $S_8  = \sigma_8 \sqrt{\om/0.3}$ from a single galaxy survey data set, demonstrating the power of such joint two-point correlation analyses.

High resolution, low-noise observations of the cosmic microwave background (CMB) have recently enabled mapping of gravitational lensing of the CMB, typically quantified via the lensing convergence, $\kappa_{\rm CMB}$.  While it is possible to convert a map of the convergence to shear, doing so is not necessary for this analysis.  Two-point functions that correlate $\kappa_{\rm CMB}$ with the $\delta_{\rm g}$ and $\gamma$ fields also contain cosmological information \citep{hand15,Liu:2016,kirk16,hd17,giannantonio16}. Jointly fitting $\gamma\kcmb$ and $\delg\kcmb$ with the \3x2pt cross-correlations serves several purposes. First, the joint fit helps improve cosmological constraints by breaking degeneracies with galaxy bias \citep[e.g.][]{Baxter:2016}. Second, the joint fit can constrain nuisance parameters associated with sources of systematic error in galaxy lensing measurements \citep[e.g.][]{Vallinotto:2012,Das:2013,Baxter:2016,Schaan:2017}. This is possible because the sources of systematic error that affect the measurement of $\kappa_{\rm CMB}$ are generally different from those impacting the measurement of $\gamma$. Finally, cross-correlations with CMB lensing include some sensitivity to the angular diameter distance to the last scattering surface, which can lead to improved cosmological sensitivity relative to cross-correlations with lower redshift lensing measurements.

The South Pole Telescope (SPT) \citep{Carlstrom:2011} and \Planck \citep{Tauber2010,Planck2011} provide high signal-to-noise maps of the CMB overlapping with the DES survey, allowing for the joint measurement of all six of the two-point functions that can be formed from $\delta_{\rm g}$, $\gamma$ and $\kappa_{\rm CMB}$. We will refer to the combination of all six two-point functions as \6x2pt, and the combination of all two-point functions except for $\kcmb\kcmb$ correlation as \5x2pt.  

In this work, we develop the methodology for jointly analyzing the \5x2pt set of correlation functions.  This methodology will be applied to measurements of the \5x2pt two-point functions using data from DES, SPT and \Planck in a companion paper, extending the \3x2pt analysis of \citet{DESy1:2017}.   We do not include $\kk$ in the analysis presented here because the current highest signal-to-noise measurement of $\kk$ comes from \citet{Planck:cmblensing}. Since the \Planck $\kappa_{\rm CMB}$ map covers the full sky, the covariance between $\kk$ measured by \Planck and set of \5x2pt correlations involving current SPT and DES Y1 data (which overlap over roughly 1300 sq. deg. on the sky) is negligible. Therefore, cosmological constraints from the \Planck measurement of $\kk$ can be trivially combined with those from the \5x2pt analysis by taking the product of the corresponding posteriors. For future DES and SPT data, the improved signal-to-noise of the measurements may necessitate revisiting the approximation of negligible covariance between the \Planck measurement of $\kk$ and the DES and SPT measurements of \5x2pt.

The analysis presented here builds on the methodology presented in \citet{Krause:2017} (hereafter \citetalias{Krause:2017}) for analyzing the \3x2pt data vector.  The most significant difference between this work and that of \citetalias{Krause:2017} is that we must account for sources of systematic error that are specific to the cross-correlations with $\kappa_{\rm CMB}$. Of these systematics, the most problematic is contamination of  $\kappa_{\rm CMB}$ by the thermal Sunyaev-Zel'dovich effect (tSZ).  The effects of tSZ and other potential contaminants on $\kappa_{\rm CMB}$ has been investigated previously by several authors, including \citet{vanEngelen:2014,Ferraro:2018,Madhavacheril:2018}.  We develop an approach for estimating the effects of such contamination on $\galk$ and $\sheark$, and use these estimates to determine an appropriate choice of angular scale cuts to apply to the two-point function measurements to minimize tSZ-induced bias.

After developing the methodology for analyzing the \5x2pt data vector, we use simulated likelihood analyses to demonstrate how adding the cross-correlations with $\kappa_{\rm CMB}$ to the \3x2pt analysis can improve cosmological constraints and can potentially allow for the self-calibration of nuisance parameters that are degenerate with cosmology in the \3x2pt analysis.   While the currently low signal-to-noise of the $\galk$ and $\sheark$ correlation functions limits their cosmological constraining power, we show that including them in the joint analysis can make the cosmological constraints more robust to multiplicative shear biases.

This work builds on several recent DES collaboration papers that analyze two-point functions of DES observables.  These include the analysis of cosmic shear \citep{Troxel:2017}, the analysis of galaxy clustering \citep{ElvinPoole2017}, the analysis of galaxy-galaxy lensing \citep{Prat2017}, and the joint analysis of all three two-point functions in \citet{DESy1:2017}. 

The layout of the paper is as follows. In \S\ref{sec:data} we describe the datasets used in this work; in \S\ref{sec:modeling} we describe the modeling steps required to compute a likelihood for the observed two-point functions given a cosmological model; in \S\ref{sec:kappa_systematics} we describe our procedure for characterizing systematic biases in $\galk$ and $\sheark$ that are specific to the $\kappa_{\rm CMB}$ maps; in \S\ref{sec:scale_cuts} we describe the motivation for our choice of angular scale cuts. We present results from simulated analyses in \S\ref{sec:forecasts} and conclude in \S\ref{sec:conclusion}. 

\section{Data}
\label{sec:data}

This work presents the {\it methodology} for analyzing the two-point functions formed between $\delta_{\rm g}$, $\gamma$ and $\kappa_{\rm CMB}$.  For the most part, developing this methodology does not rely on analyzing any actual data.  However, in \S\ref{sec:kappa_systematics}, we will take a data-driven approach to characterizing biases in $\galk$ and $\sheark$ due to contamination of the $\kappa_{\rm CMB}$ maps.  For that part of the analysis, we rely on exactly the same galaxy and shear catalogs used in the DES \3x2pt analysis \citep{DESy1:2017}.  Below, we briefly describe these catalogs and refer readers to the listed references for more details.

We consider measurements of $\galk$ and $\sheark$ in position-space, i.e. as a function of the angle between the two points being correlated.  Measuring $\galk$ and $\sheark$ requires two sets of galaxies, which we refer to as `tracers' and `sources.'  Lenses are treated as tracers of the matter density field and are used to measure $\delta_{\rm g}$; images of the source galaxies are used to measure the gravitational lensing-induced shears, $\gamma$.  The tracer and source galaxies are in turn divided into multiple redshift bins.

\subsection{Galaxy catalog}
\label{sec:redmagic}

For the purposes of measuring $\delta_{\rm g}$, we use a subset of the DES Y1 `Gold' catalog \citep{Drlica-Wagner17} referred to as redMaGiC \citep{Rozo2016redmagic}.  The redMaGiC galaxies are a set of luminous red galaxies (LRGs) selected based on their match to a red sequence template, which is calibrated via the redMaPPer galaxy-cluster-finding algorithm \citep{Rykoff2014,Rozo2016redmagic, Rykoff2016}.  The redMaGiC galaxies are designed to have very well understood photometric redshift estimates, with a scatter of $\sigma_z \sim 0.017 (1+z)$ \citep{ElvinPoole2017}.  As in \citetalias{Krause:2017}, the redMaGiC galaxies are divided into 5 redshift bins at $0.15 \lesssim z \lesssim 0.9$, where the three lower redshift bins have a luminosity threshold of $L_{\min}=0.5L^{*}$ and the two higher redshift bins have luminosity thresholds $L_{\min}=1.0 L^{*}$ and $1.5L^{*}$, respectively. For a more detailed description of the galaxy sample, see also \cite{Prat2017} and \cite{ElvinPoole2017}.

\subsection{Shear catalog}
\label{sec:shear}

For the purposes of measuring $\gamma$, we use the same shear catalogs used in the \3x2pt analysis. Two shear measurement algorithms -- \textsc{MetaCalibration} \citep{Huff17,sheldon17} and \textsc{Im3shape} \citep{zuntz13} -- were used to generate the galaxy shear catalogs that were used in the \3x2pt analysis, while the \textsc{MetaCalibration} catalog was used as the fiducial catalog due to its higher signal-to-noise. \textsc{MetaCalibration} uses the data itself to calibrate the bias in shear estimation by artificially shearing the galaxy images and re-measuring the shear. \textsc{Im3shape}, on the other hand, invokes a large number of sophisticated image simulations to calibrate the bias in shear estimates.  As in \citetalias{Krause:2017}, the shear catalogs were divided into four redshift bins between $z\sim 0.2$ and 1.3.  For a detailed description of both shear catalogs, see \citet{Zuntz17}.  For details of the photo-z catalog associated with the shear catalogs, see \citet{Hoyle2017}.  The analysis presented in this work adopts noise estimates and redshift distributions corresponding to the \textsc{MetaCalibration} catalog.

\subsection{CMB lensing map}

The methodology presented here is general and could be applied to any map of $\kappa_{\rm CMB}$. However, in order to accurately characterize the magnitude of biases in $\kappa_{\rm CMB}$, we tailor our analysis to the $\kappa_{\rm CMB}$ maps that will be used in the companion paper that presents cosmological constraints obtained from analysis of the \5x2pt data vector.  That work will use the $\kappa_{\rm CMB}$ maps from \citet{Omori:2017} (henceforth \citetalias{Omori:2017}) and so we briefly describe those maps here.

The $\kappa_{\rm CMB}$ map generated in \citetalias{Omori:2017} is computed by applying the quadratic lensing estimator of \citet{Hu:2002} to an inverse variance weighted combination of 150 GHz SPT and 143 GHz \Planck temperature maps.  The quadratic estimator of \citet{Hu:2002} exploits the fact that gravitational lensing induces a correlation between the gradient of the CMB temperature field and small-scale fluctuations in this field.  A suitably normalized quadratic combination of filtered CMB temperature maps then provides an estimate of $\kappa_{\rm CMB}$.  The SPT maps used for this purpose are from the SPT-SZ survey \citep{Story:2013}.  The combined map produced from the SPT and \Planck datasets is sensitive to a greater range of angular modes on the sky than either experiment alone:  \Planck cannot measure small scale modes because of its 7' beam (at 143 GHz), while SPT cannot measure large scale modes because of time domain filtering that is used to remove atmospheric contamination.  

The $\kappa_{\rm CMB}$ map from \citet{Omori:2017} is restricted to the area of sky that is observed by both SPT and \Planck.  The overlap of this region with the DES Y1 survey region is approximately 1300 sq. deg.

\section{Modeling the two-point functions}
\label{sec:modeling}

\subsection{Formalism}
\label{sec:formalism}

We begin by describing the formalism used to model the \5x2pt set of correlation functions. This methodology closely follows that described in \citetalias{Krause:2017} to model the \3x2pt data vector. We consider exactly the same galaxy selections, and make many of the same modeling assumptions. To minimize repetition, in this work, we focus only on describing the modeling of those correlations that involve $\kappa_{\rm CMB}$ (i.e. $\galk$ and $\sheark$); for a complete description of the modeling of the other two-point functions (i.e. $\galgal$, $\galshear$, and $\shearshear$), we refer readers to \citetalias{Krause:2017}.  

Since shear defines a spin-2 field, we can consider correlations with different components of this field.  When considering autocorrelations of the shear field, we use $\xi_{+}$ and $\xi_{-}$ \citep{Bartelmann2001}.  When measuring the correlation between DES shears and $\kappa_{\rm CMB}$, we consider only the component of the shear that is oriented orthogonally to the line connecting the two points being correlated, i.e. the tangential shear, $\gamma_{\rm t}$.  In the weak shear limit, this tangential component contains all the lensing signal \citep{Bartelmann2001}. Using $\gamma_{\rm t}$ has the advantage of reducing contamination from additive systematics in the shear estimation and avoiding mask effects during the conversion from $\gamma$ to $\kappa$ \citep{hd16}. Henceforth, we will denote this correlation as $\sheartk$.

We begin by computing the cross-spectra between the relevant fields in harmonic space using the Limber approximation \citep{Limber:1953}.  The Limber approximation is justified here because we do not consider very large angular scales, and because the galaxy selection functions are slowly varying with redshift \citep{Loverde:2008}.  For computing $\sheartk$, it is convenient to first express this cross-correlation in terms of lensing convergence, rather than shear, and then transform to shear when expressing the correlation function in configuration space.  The lensing convergence, $\kappa$, in some direction specified by $\hat{\theta}$, is defined by
\begin{equation}
\kappa(\hat{\theta}, \chi_{\rm s}) = \frac{3\om H_0^2}{2c^2} \int_{0}^{\chi_{\rm s}} d\chi' \frac{\chi'(\chi - \chi')}{\chi} \frac{\delta(\hat{\theta},\chi')}{a(\chi')},
\end{equation}
where $\chi$ is comoving distance (with $\chi_s$ being the comoving distance to the source plane), $H_{0}$ is the Hubble constant today, $\om$ is the matter density today, $\delta$ is the matter overdensity, and $a$ is the scale factor \citep{Bartelmann2001}.  We refer to the lensing convergence defined for the source galaxies as  $\kappa_{\rm s}$ (in contrast to the CMB-derived lensing convergence, $\kappa_{\rm CMB}$).  For galaxy lensing, the sources are distributed across a broad range of redshift and the convergence must be averaged across this distribution.  In this case, the convergence for source galaxies in the $i$th redshift bin becomes
\begin{equation}
\kappa_{\rm s}^i(\hat{\theta}) = \int_0^{\infty} d\chi' q_{\kappa_{\rm s}}^i(\chi') \delta(\hat{\theta},\chi'),
\end{equation}
where we have defined the lensing weight as
\begin{equation}
q_{\kappa_{\rm s}}^i (\chi) = \frac{3 \Omega_{\rm m} H_0^2}{2c^2}\frac{\chi}{a(\chi)}\int_{\chi}^{\infty} d\chi' \frac{n^i_{\rm s}(z(\chi')) \frac{dz}{d\chi'}}{\bar{n}_{\rm s}^i} \frac{\chi' - \chi}{\chi'},
\label{eq:weight_kappa}
\end{equation}
where $n^i_{\rm s}(z)$ the number density of the source galaxies in the $i$th bin as a function of redshift, and $\bar{n}_{\rm s}^i$ is the average of that quantity over redshift.  Since the CMB originates from a very narrow range of comoving distance, we can approximate the source redshift distribution of the CMB as a Dirac $\delta$-function centered on the comoving distance to the last scattering surface, $\chi^*$.  In this case, the lensing weight function for CMB lensing becomes
\begin{equation}
q_{\kappa_{\rm CMB}} (\chi) = \frac{3\om H_0^2}{2c^2}\frac{\chi}{a(\chi)}  \frac{\chi^* - \chi}{\chi^*}. \label{eq:weight_cmbkappa}
\end{equation}

The overdensity of galaxies on the sky in the $i$th redshift bin can also be related to an integral along the line of sight of the matter overdensity, assuming the galaxy bias is known.  Following \citet{DESy1:2017}, we restrict our analysis to the linear bias regime, where the galaxy overdensity can be expressed as $\delta_{\rm g}(\hat{\theta}, \chi) = b_{\rm g}(\chi) \delta(\hat{\theta}, \chi)$, where $b_{\rm g}(\chi)$ is the galaxy bias.  In this case, the projected overdensity of galaxies on the sky is 
\begin{equation}
\delta_{\rm g}^i (\hat{\theta}) = \int d\chi' q_{\delta_{\rm g}}^i(\chi') \delta(\hat{\theta}, \chi'),
\end{equation}
where we have defined the tracer galaxy weight function as
\begin{equation}
q_{\delta_{\rm g}}^i (\chi) = b_{\rm g}^i(\chi) \frac{n_{\rm g}^i(z(\chi))}{\bar{n}_{\rm g}^i} \frac{dz}{d\chi}, \label{eq:weight_gal}
\end{equation}
where $n_{\rm g}^i(z)$ is the number density of the tracer galaxies in the $i$th bin as a function of redshift, and $\bar{n}_{\rm g}^i$ is the average of that quantity over redshift.  We will further simplify the bias modeling such that the bias for each galaxy redshift bin is assumed to be a constant, $b_{\rm g}^i$.  In reality, the linear bias model is known to break down at small scales \citep{Zehavi2005,Blanton2006,Cresswell2009}.  We will show in \S\ref{sec:scale_cuts} that for our choice of angular scale cuts, the assumption of linear bias does not bias our parameter constraints.

Using the Limber approximation, we have
\begin{equation}
\kappaskell = \\\int d\chi \frac{q^i_{\kappa_{\rm s}} (\chi) q_{\kappa_{\rm CMB}} (\chi) }{\chi^2} P_{\rm NL} \left( \frac{\ell+1/2}{\chi}, z(\chi)\right),
\label{eq:CGK}
\end{equation}
and
\begin{equation}
\galkell = \\
\int d\chi \frac{q^i_{\delta_{\rm g}} \left(\chi \right) q_{\kappa_{\rm CMB}} (\chi) }{\chi^2} P_{\rm NL} \left( \frac{\ell+1/2}{\chi}, z(\chi)\right),
\label{eq:CNK}
\end{equation}
where $i$ labels the redshift bin (of either the tracer or source galaxies) and $P_{\rm NL}(k,z)$ is the nonlinear matter power spectrum. We compute the nonlinear power spectrum using the Boltzmann code CAMB\footnote{See \texttt{camb.info}.} \citep{Lewis:2000,Howlett:2012} with the Halofit extension to nonlinear scales \citep{Smith:2003,Takahashi:2012} and the \citet{Bird:2002} neutrino extension.

\begin{figure*}
\begin{center}
\includegraphics[width=1.0\textwidth]{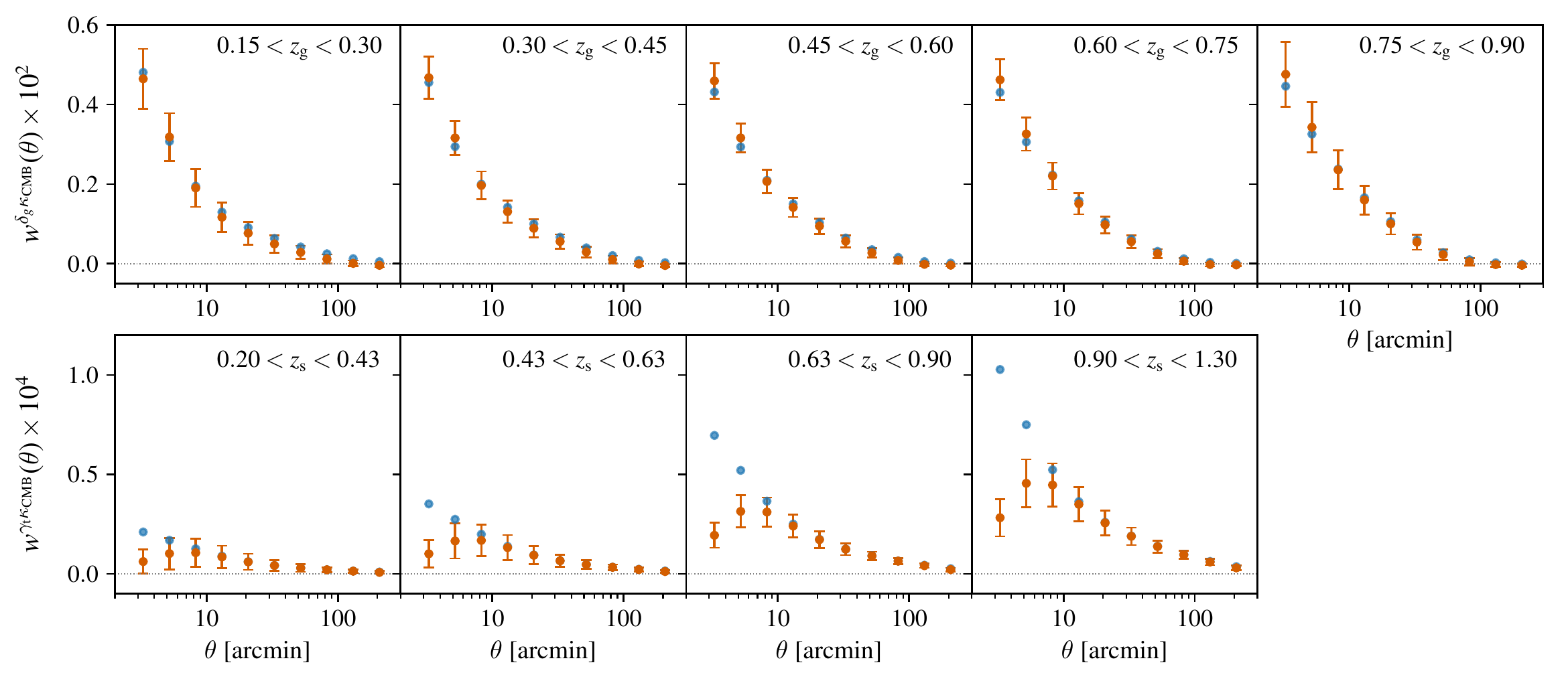}
\caption{Models of the $\galk$ and $\sheartk$ correlation functions corresponding to the fiducial cosmological model of Table~\ref{tab:params} (orange points with errorbars). Each panel represents the correlation function for a different tracer or source redshift bin.  Error bars correspond to the square root of the diagonal elements of the covariance matrix described in \S\ref{sec:covariance}.  Blue points show the model vectors in the absence of the harmonic-space filtering of the $\kappa_{\rm CMB}$ map described in \S\ref{sec:formalism}; the filtering affects $\galk$ and $\sheartk$ differently because of the non-local nature of $\gamma_{\rm t}$.}
\label{fig:models}
\end{center}
\end{figure*}

SPT and \Planck observe the CMB with finite-size beams.  When generating the $\kappa_{\rm CMB}$ map, this beam is deconvolved, exponentially increasing noise at small scales.  Unfortunately, the presence of small-scale noise in $\kappa_{\rm CMB}$ will make the real-space covariance diverge.  To prevent this divergence, we apply a smoothing function to the $\kappa_{\rm CMB}$ maps.  We convolve the maps with a Gaussian beam having full width at half maximum of $\theta_{\rm FWHM} = 5.4'$.  In harmonic space, this corresponds to multiplication of the maps by
\begin{equation}
B(\ell) = \exp (-\ell(\ell + 1)/\ell_{\rm beam}^2),
\end{equation}
where $\ell_{\rm beam} \equiv \sqrt{16 \ln 2}/\theta_{\rm FWHM} \approx 2120$.  Additionally, we filter out modes in the $\kappa_{\rm CMB}$ map with $\ell < 30$ and $\ell > 3000$, where the lower bound is to remove biases coming from poorly characterized modes due to the finite sky area covered by the $\kcmb$ lensing map \citep{Omori:2017} and the upper limit is imposed to remove potential biases due to foregrounds in the $\kappa_{\rm CMB}$ map.  The impact of this filtering can be seen in Fig.~\ref{fig:models}.

Converting the above expressions to position-space correlation functions via a Legendre transform yields
\begin{eqnarray}
\sheartik &=& \int \frac{d\ell \, \ell}{2\pi} F(\ell) J_2(\ell \theta) \kappasikell,\label{eq:shearkcorr} \\
\galik &=& \sum \frac{2\ell + 1}{4\pi} F(\ell)P_{\ell} (\cos(\theta)) \galikell,
\end{eqnarray}
where $J_2$ is the second order Bessel function of the first kind and $P_{\ell}$ is the $\ell$th order Legendre polynomial.  The appearance of $J_2$ in Eq.~\ref{eq:shearkcorr} is a consequence of our decision to measure the correlation of $\kappa_{\rm CMB}$ with tangential shear.  The function $F(\ell) = B(\ell) \Theta(\ell - 30) \Theta(3000 - \ell)$, where $\Theta(\ell)$ is a step function, describes the filtering that is applied to the $\kappa_{\rm CMB}$ map.  Henceforth, for notational convenience, we will suppress the redshift bin labels on the correlation functions.  We show the model $\galk$ and $\sheartk$ corresponding to the best-fit \Planck cosmological parameters in Fig.~\ref{fig:models}.

\subsection{Modeling systematics affecting $\delta_{\rm g}$ and $\gamma$}
\label{sec:3x2_systematics}

There are several sources of systematic uncertainty that affect the $\delta_{\rm g}$ and $\gamma$ observables.  These systematics will propagate into the $\galk$ and $\sheartk$ measurements.  We model these sources of systematic error exactly as described in \citetalias{Krause:2017}, and so provide only a brief description here.  We will consider sources of systematic error that can affect the $\kappa_{\rm CMB}$ map in more detail in \S\ref{sec:kappa_systematics}.  

\subsubsection{Shear calibration bias}
The inference of $\gamma$ from an image of a galaxy is  subject to sources of systematic error.  Such errors are commonly parameterized in terms of a multiplicative bias, $m$, such that the observed shear is related to the true shear by $\gamma_{\rm obs} = (1+m)\gamma_{\rm true}$ \citep[e.g.][]{Zuntz17}.  While additive biases may also be present in shear calibration, these are typically tightly constrained by the data itself (and are minimized by our decision to use the tangential shear component).

Following \citetalias{Krause:2017} and other literature \citep{Abbott2015,Joudaki2017,Hildebrandt2017}, we adopt a separate multiplicative bias parameter, $m_i$, for the $i$th source galaxy redshift bin.  The model for $\sheartk$ (Eq.~\ref{eq:shearkcorr}) is then scaled by $(1+m_i)$.  Note, however, that $\galk$ does not depend on the estimated shears and is therefore unaffected by $m_i$.

\subsubsection{Intrinsic alignment}

In addition to the coherent alignment of galaxy shapes caused by gravitational lensing, galaxy shapes can also be intrinsically aligned as a result of e.g. tidal fields \citep{Heavens2000,Catelan2001,Crittenden2001}.  Such  {\it intrinsic alignments} constitute a potential systematic for the measurement of gravitational lensing from galaxy shapes.  Intrinsic alignments of galaxies will also affect $\sheartk$ \citep{Hall:2014,Troxel:2014}.  To see this, consider a galaxy that is stretched by the tidal field of nearby large scale structure; the same large scale structure that causes this intrinsic alignment will also lens the CMB, leading to a correlation between the intrinsic galaxy shapes and $\kappa_{\rm CMB}$. This effect is analogous to the usual gravitational-intrinsic (GI) term affecting cosmic shear \citep{Hirata:2004}.  Following \citetalias{Krause:2017}, we parameterize the effects of intrinsic alignments using the nonlinear linear alignment (NLA) model \citep{bridle07}.  This model impacts $q_{\kappa}$ for the source galaxies as described in \citetalias{Krause:2017}.  

Briefly, we perform the replacement
\begin{eqnarray}
q_{\rm \kappa_{\rm s}}^i(\chi) \rightarrow q_{\kappa_{\rm s}}^i(\chi) - A(z(\chi)) \frac{n_{\rm s}^i(z(\chi))}{\bar{n}_{\rm s}^i} \frac{dz}{d\chi},
\end{eqnarray}
where
\begin{eqnarray}
A(z) = A_{{\rm IA}, 0} \left( \frac{1+z}{1+z_0} \right)^{\eta_{\rm IA}} \frac{0.0139\Omega_{\rm m} }{D(z)},
\end{eqnarray}
where $D(z)$ is the linear growth factor and we set $z_0 = 0.62$.  The normalization $A_{\rm IA,0}$ and power law scaling with redshift, $\eta_{\rm IA}$ are treated as free parameters of the model.

\subsubsection{Photometric redshift errors}
DES uses multiband optical photometry to infer the redshift distributions of the galaxy samples (it is these distributions that are necessary for modeling the \5x2pt set of correlation functions).  This inference is potentially subject to sources of systematic error, which can result in biases to $n_{\rm g}(z)$ and $n_{\rm s}(z)$.  Following \citetalias{Krause:2017} and other literature \citep{Abbott2015,Joudaki2017,Hildebrandt2017}, we parameterize such biases in terms of the shift parameters, $\Delta_{\rm z}$, such that the estimated redshift distribution, $\hat{n}(z)$ is related to the true redshift distribution, $n_{\rm true}(z)$, via $n_{\rm true}(z) = \hat{n}(z - \Delta_{z})$.  We consider separate shift parameters for each tracer and source galaxy redshift bin, $\Delta_{z,{\rm g}}^i$ and $\Delta_{z,{\rm s}}^i$, respectively, where the $i$ superscript labels the redshift bin. 

\subsection{Covariance}
\label{sec:covariance}

The DES \3x2pt analysis uses a halo model covariance, as described and validated in \citetalias{Krause:2017}. We now describe the extension of this formalism to the CMB lensing cross-correlations $\galk$ and $\sheartk$.  For notational convenience, we will use $\Sigma(\theta)$ and $\Sigma(\ell)$ to generically represent one of these two-point functions in position and harmonic-space, respectively; we will use $\Xi(\theta)$ and $\Xi(\ell)$ to represent one of the \3x2pt correlation functions (i.e. $\wtheta$, $\galsheart$, $\xi_{+}(\theta)$ and $\xi_{-}(\theta)$) in configuration and harmonic space, respectively.  We calculate the covariance of the harmonic space correlation functions, $\mathrm{Cov}(\Sigma^i(\ell), \Sigma^k(\ell'))$ as the sum of a Gaussian covariance $\mathrm{Cov}^{\mathrm{G}}$ and non-Gaussian covariance $\mathrm{Cov}^{\mathrm{NG}}$, which includes super-sample variance \citep{Takada:13}, as detailed in \citet{cosmolike} and \citet{Schaan:2017}, using the halo model to compute the higher-order matter correlation functions.  The covariance of the $\galk$ and $\sheartk$ is then
\begin{multline}
\mathrm{Cov}\left(\Sigma^{i}(\theta),\, \Sigma^{k}(\theta')\right) = \int \frac{d\ell\,\ell}{2\pi} J_{n(\Sigma^{i})}(\ell \theta) F(\ell)\int \frac{d\ell'\,\ell'}{2\pi}  J_{n(\Sigma^{k})}(\ell' \theta') F(\ell') \\
\left[\mathrm{Cov}^{\mathrm{G}}\left(\Sigma^{i}(\ell),\Sigma^{k}(\ell')\right)+\mathrm{Cov}^{\mathrm{NG}}\left(\Sigma^{i}(\ell),\Sigma^{k}(\ell')\right)\right]\,,
\label{eq:FT_cov}
\end{multline}
where $J_{n}$ is the $n$th-order Bessel function of the first kind, and $F(\ell)$ is the function that describes the filtering that is applied to the $\kappa_{\rm CMB}$ map. The cross-covariance between $\galk$ and $\sheartk$ with one of the DES \3x2pt correlation functions is given by
\begin{multline}
\mathrm{Cov}\left(\Sigma^{i}(\theta),\, \Xi^{k}(\theta')\right) = \int \frac{d\ell\,\ell}{2\pi} J_{n(\Sigma)}(\ell \theta) \int \frac{d\ell'\,\ell'}{2\pi}  J_{n(\Xi)}(\ell' \theta') \\
\left[\mathrm{Cov}^{\mathrm{G}}\left(\Sigma^{i}(\ell),\Xi^{k}(\ell') \right) + \mathrm{Cov}^{\mathrm{NG}}\left(\Sigma^{i}(\ell),\Xi^{k}(\ell')\right)\right]\,,
\label{eq:FT_cov2}
\end{multline}
where the order of the Bessel function is given by $n = 0$ for $\galk$, $\wtheta$, and $\xi_{+}$, by $n = 2$ for $\sheartk$ and $\galsheart$, and by $n = 4$ for $\xi_{-}$. 

\renewcommand{\arraystretch}{1.3}
\begin{table}
\caption{Parameters of the baseline model: fiducial values, flat priors (min, max), and Gaussian priors ($\mu$, $\sigma$).  Definitions of the parameters can be found in the text.  The cosmological model considered is spatially flat $\Lambda$CDM, so the curvature density parameter and equation of state of dark energy are fixed to $\Omega_{\mathrm K} = 0$ and $w=-1$, respectively.}
\begin{center}
\begin{tabular*}{0.45\textwidth}{@{\extracolsep{\fill}}| c c c |}
\hline
\hline
Parameter & Fiducial & Prior \\  
\hline 
\multicolumn{3}{|c|}{\textbf{Cosmology}} \\
$\om$ & 0.295 &  flat (0.1, 0.9)  \\ 
$A_\mathrm{s}/10^{-9}$ &$2.26$ &  flat ($0.5$,$5.0$)  \\ 
$n_{\rm s}$ & 0.968 & flat (0.87, 1.07)  \\
$\w$ &  -1.0 &   fixed   \\
$\omb$ &  0.0468 &  flat (0.03, 0.07)  \\
$h_0$  & 0.6881 &  flat (0.55, 0.91)   \\
$\Omega_\nu h^2$ & $6.16\times 10^{-4}$ & fixed \\
$\Omega_\mathrm{K}$ & $0$ & fixed \\
\hline
\multicolumn{3}{|c|}{\textbf{Galaxy bias}} \\
$b_\mr{g}^1$ & 1.45  & flat (0.8, 3.0) \\
$b_\mr{g}^2$ & 1.55  &flat (0.8, 3.0) \\
$b_\mr{g}^3$ & 1.65 & flat (0.8, 3.0) \\
$b_\mr{g}^4$ & 1.8 & flat (0.8, 3.0) \\
$b_\mr{g}^5$ & 2.0 & flat (0.8, 3.0) \\
\hline
\multicolumn{3}{|c|}{\textbf{Tracer galaxy photo-z bias}} \\
$\Delta_{z,{\rm g}}^1 $ & 0.002 & Gauss (0.0, 0.007) \\
$\Delta_{z,{\rm g}}^2 $ & 0.001 & Gauss (0.0, 0.007) \\
$\Delta_{z,{\rm g}}^3 $ & 0.003 & Gauss (0.0, 0.006) \\
$\Delta_{z,{\rm g}}^4 $ & 0.0 & Gauss (0.0, 0.01) \\
$\Delta_{z,{\rm g}}^5 $ & 0.0 & Gauss (0.0, 0.01) \\
\hline
\multicolumn{3}{|c|}{\textbf{Source galaxy photo-z bias}} \\
$\Delta_{z,{\rm s}}^1 $ & -0.002 & Gauss (-0.001,0.016) \\
$\Delta_{z,{\rm s}}^2 $ & -0.0015 & Gauss (-0.019,0.013) \\
$\Delta_{z,{\rm s}}^3 $ & 0.007 & Gauss (0.009, 0.011) \\
$\Delta_{z,{\rm s}}^4 $ & -0.018 & Gauss (-0.018, 0.022) \\
\hline
\multicolumn{3}{|c|}{\textbf{Shear Calibration bias }} \\
$m^i $ & 0.013 & Gauss (0.012, 0.023)\\
\hline
\multicolumn{3}{|c|}{\textbf{Intrinsic Alignments}} \\
$A_{\mathrm{IA,0}} $ & 0.0 & flat (-5.0, 5.0)\\
$\eta_{\mathrm{IA}} $ & 0.0 & flat (-5.0, 5.0)\\
$z_0$& 0.62 & fixed\\
\hline
\end{tabular*}
\end{center}
\label{tab:params}
\end{table}
\renewcommand{\arraystretch}{1.0}

\subsection{Likelihood analysis}
\label{sec:likelihoodanalysis}

We now build the likelihood of the data given the model described in \S\ref{sec:formalism} and the covariance described in \S\ref{sec:covariance}. The model includes parameters describing cosmology, galaxy bias, intrinsic alignment, and shear and photo-$z$ systematics.  The cosmological model considered in this analysis is flat $\Lambda$CDM.  The cosmological parameters varied are the present day matter density parameter, $\om$, the normalization of the primordial power spectrum, $A_\mathrm{s}$, the spectral index of the primordial power spectrum, $n_{\rm s}$, the present day baryon density parameter, $\omb$, and the Hubble parameter today, $h_0$.  The complete set of model parameters is summarized in Table~\ref{tab:params}.  For the simulated likelihood analyses described below, we generate a data vector at a fiducial set of model parameters given by the middle column of Table~\ref{tab:params}.  The priors imposed in our fiducial likelihood analysis are given in the third column of Table~\ref{tab:params}; these priors are identical to those of the \3x2pt analysis of \citet{DESy1:2017}.  

For the purposes of this analysis, we keep the cosmological density of neutrinos fixed to $\Omega_{\nu}h^2 = 6.16\times10^{-4
}$, corresponding to a total neutrino mass of 0.06 eV.  This choice is reasonable since the \citet{DESy1:2017} analysis only weakly constrains the neutrino mass, and the \5x2pt analysis does not significantly improve on these constraints.  

Given a point in parameter space, $\mathbf{p}$, we consider a Gaussian likelihood for the \5x2pt observable, $\mathbf{d}$:
\begin{multline}
\mathcal{L}(\mathbf{d}|\mathbf{p}) \propto \exp \left[ -\frac{1}{2}\sum_{ij} (d_i - m_i(\mathbf{p})) \left[ \mathbf{C}^{-1} \right]_{ij} (d_j - m_j(\mathbf{p})) \right], 
\end{multline}
where $\mathbf{m}$ is the model vector, the sum runs over all elements of the data vector, and $\mathbf{C}$ is the covariance matrix described in \S\ref{sec:covariance}.  As in \citetalias{Krause:2017}, we keep the covariance matrix fixed as a function of cosmological parameters.  This ignores the cosmology-dependence of the covariance matrix \citep{Morrison2013JCAP,Eifler2009}, which is negligible compared to the noise level in the DES Y1 and SPT data. 

The computation of the model vector and the likelihood analysis is accomplished using \texttt{CosmoSIS} \citep{Zuntz:2015}.  We sample parameter space using the \texttt{multinest} algorithm \citep{Feroz2009}. The \texttt{multinest} sampler has been tested in \citetalias{Krause:2017} to yield results consistent those of another sampler, \texttt{emcee} \citep{ForemanMackey2013}, which relies on the algorithm of \citet{Goodman:2010}.

\section{Biases in the $\kappa_{\rm CMB}$ maps}
\label{sec:kappa_systematics}

\subsection{Overview}

While the systematics considered in \S\ref{sec:3x2_systematics} affect both the \3x2pt data vector and the \5x2pt data vector, there are also sources of systematic error that impact only $\galk$ and $\sheartk$. In this section, we attempt to quantify biases in the $\kappa_{\rm CMB}$ maps that will affect the measurement of these two correlation functions.  

We write the observed $\kappa_{\rm CMB}$ signal on the sky, $\kappa_{\rm obs}$, as the sum of the true CMB lensing signal, $\kappa_{\rm CMB}$, and some contaminating field, $\kappa_{\rm sys}$, i.e. $\kappa_{\rm obs} = \kappa_{\rm CMB} + \kappa_{\rm sys}$.  The observed correlation functions $\galkobs$ and $\sheartkobs$ then differ from the correlation functions with the true $\kappa_{\rm CMB}$ by $\sheartksys$ and $\galksys$.  To determine these biases, we will form an estimate of $\kappa_{\rm sys}$ and then use the true galaxy and shear catalogs described in \S\ref{sec:data} to calculate $\sheartksys$ and $\galksys$. However, given the large uncertainties associated with our estimates of $\kappa_{\rm sys}$, we will not attempt to model or correct for such biases in our analysis.  Instead, we will choose angular scale cuts such that biases to the inferred posteriors on the model parameters are below 50\% of the statistical errors (see discussion in \S\ref{sec:scale_cuts}).

The dominant sources of bias that contribute to $\kappa_{\rm sys}$ will depend on the methods and data used to estimate $\kappa_{\rm CMB}$.  For instance, a $\kappa_{\rm CMB}$ map created from maps of CMB temperature will be affected by the tSZ effect, while this is not the case for $\kappa_{\rm CMB}$ maps constructed from maps of CMB polarization.  Here we tailor our analysis to those systematics that are expected to be dominant for the cross-correlation of DES galaxies and shears with the $\kappa_{\rm CMB}$ maps generated in \citetalias{Omori:2017}, since it is these $\kappa_{\rm CMB}$ maps that will be used in the forthcoming \5x2pt results paper. 

Both the SPT 150 GHz maps and \Planck 143 GHz maps used to construct the $\kappa_{\rm CMB}$ maps in \citetalias{Omori:2017} receive contributions from sources other than primary CMB.  In particular, these maps receive significant contributions from the tSZ effect and from radio and thermal dust emission from distant galaxies.  The tSZ effect is caused by inverse Compton scattering of CMB photons with hot electrons.  At frequencies near 150 GHz, this results in a decrement in the observed CMB temperature. Unresolved galaxies, which together constitute the cosmic infrared background (CIB), on the other hand, appear as a diffuse background in the observed maps.  The tSZ and CIB signals on the sky will propagate through the quadratic estimator into the $\kappa_{\rm CMB}$ maps of \citetalias{Omori:2017}.  Since both non-Gaussian sources of contamination are correlated with the matter density, we also expect $\kappa_{\rm sys}$ to be correlated with the matter density.  Consequently, these biases will not average to zero in the $\galk$ and $\sheartk$ correlations, and we must carefully quantify their impact on our analysis.  Note that contamination from the kinematic Sunyaev-Zel'dovich (kSZ) effect is also expected to be present in the $\kappa_{\rm CMB}$ maps.  However, since the kSZ signal has a similar morphology to the tSZ signal, but an amplitude that is a factor of $\sim 10$ smaller, by ensuring that the tSZ effect does not bias our results, we ensure that the kSZ effect also does not lead to a significant bias.

Our approach to estimating $\kappa_{\rm sys}$ due to both tSZ and CIB is to estimate the contributions to the SPT+\Planck temperature maps from these signals, and to then pass these estimated temperature maps through the quadratic estimator pipeline of \citetalias{Omori:2017}.  To see that this procedure works, consider the total temperature at some multipole, $\mathbf{\ell}$, as the sum of the lensed CMB and the contaminating signal: $T_{\rm tot}(\mathbf{\ell}) = T_{\rm CMB}(\mathbf{\ell}) + T_{\rm sys}(\mathbf{\ell})$.  The quadratic estimator for the lensing potential $\phi(\mathbf{L})$ is then $\phi(\mathbf{L}) \propto \langle (T(\mathbf{\ell}) + T_{\rm sys}(\mathbf{\ell}))(T(\mathbf{\ell}') + T_{\rm sys}(\mathbf{\ell}')) \rangle $, where $\mathbf{L} = \mathbf{\ell} + \mathbf{\ell}'$.  Under the gradient approximation, $T(\mathbf{\ell}) \approx \tilde{T}(\mathbf{\ell}) + (\nabla \tilde{T} \cdot \nabla \phi)(\mathbf{\ell})$, where the tilde denotes the unlensed field.  In the case of both tSZ and CIB bias, terms of the form $T(\mathbf{\ell}) T_{\rm sys}(\mathbf{\ell}')$  average to zero because the unlensed gradient field is uncorrelated with these biases.  Therefore, we have $\phi(\mathbf{L}) \propto \phi(\mathbf{L}) + \phi_{\rm sys} (\mathbf{L})$, where $\phi_{\rm sys}(\mathbf{L})$ is the "lensing" potential associated with the contaminating temperature field.  

As we will see below, biases in $\galk$ and $\sheartk$ due to the tSZ effect can be quite large, and dominate over all other biases considered.  Since massive galaxy clusters are the largest contributors to the tSZ effect on the sky, the level of tSZ bias in the $\kappa_{\rm CMB}$ maps can be reduced by masking these objects.   Indeed, \citetalias{Omori:2017} masked clusters detected in the SPT maps at high significance via their tSZ decrement before applying the quadratic estimator to the SPT+\Planck temperature maps.   Although masking regions of high tSZ signal reduces the tSZ-induced bias, it has the undesirable consequence of inducing {\it another} bias in the correlation functions, since the regions of high tSZ signal are also regions of high $\kappa_{\rm CMB}$.  We will argue below that this bias is negligible given our masking choices.

We emphasize that the approach taken in this section to characterizing biases in the $\kappa_{\rm CMB}$ map is quite general, and could be applied to characterize biases present in maps other than that of \citetalias{Omori:2017}.  However, the values of the biases obtained here (in particular the measurement of bias due to tSZ contamination) apply only to the $\kappa_{\rm CMB}$ maps of \citet{Omori:2017}.  Maps of $\kappa_{\rm CMB}$ generated from other data sets or using different techniques could have significantly different levels of bias.

\subsection{Estimate of bias due to the tSZ effect}
\label{sec:tszbias}

\subsubsection{Construction of simulated $y$ map}

As described above, we estimate the tSZ-induced bias in $\galk$ and $\sheartk$ by correlating the true galaxy and shear catalogs with an estimate of the bias in the $\kappa_{\rm CMB}$ map due to tSZ signal, which we refer to as $\kappa_{\rm tSZ}$.  We estimate $\kappa_{\rm tSZ}$ by applying the quadratic lensing estimator to an estimated map of the tSZ temperature signal in the SPT+\Planck sky maps.  In principle, the tSZ temperature signal could be computed directly from the multi-frequency SPT and \Planck sky maps.  Instead, we take the approach of constructing a {\it simulated} map of the tSZ signal by placing mock tSZ profiles at the locations of massive galaxy clusters on the sky.   One advantage of using a simulated tSZ map instead of generating one from SPT or \Planck temperature maps is that the simulated map will not be affected by noise in the temperature maps, making it possible to characterize the bias with high statistical accuracy.  On the other hand, this approach carries some associated modeling uncertainty, which we will attempt to constrain below.

The cluster sample used to generate the simulated tSZ map combines the redMaPPer \citep{Rykoff2014} cluster catalog from DES Y1 data with samples of tSZ-detected clusters from SPT and \Planckc.  We use redMaPPer clusters with richness $\lambda > 20$, SPT clusters with detection significance $\xi > 4.5$ \citep{Bleem:2015} and the entire \Planck tSZ-detected cluster sample \citep{Planck:clustercatalog}.  Each of these samples probes a different range of mass and redshift.  The redMaPPer sample captures low mass clusters, but only over the redshift range of DES.  The SPT cluster sample captures only very massive clusters, but out to high redshift.  The \Planck cluster sample, on the other hand, captures very massive clusters at low redshift which are missed by both SPT and DES.  

Of course, there are halos in the Universe that are not detected by redMaPPer, SPT or \Planck --- and are therefore missing from the simulated tSZ map --- but nonetheless contribute to the tSZ signal on the sky.  However, halos outside of the DES survey region or at redshifts beyond those probed by DES, will not correlate with DES galaxies and shears, and will therefore not bias the inferred correlation functions (although this tSZ contribution will contribute as noise to the measurements).  There are also halos within the DES survey region and redshift range that are not detected by any of these three surveys because their corresponding observables are below the detection limit.  The lowest mass halos in our sample come from the redMaPPer catalog.  The limiting richness threshold of the redMaPPer catalog that we employ is $\lambda = 20$, corresponding roughly to a mass of $M \sim 1.5\times 10^{14}\,M_{\odot}$ assuming the mass-richness relation of \citet{Melchior:2017}.  Using simulations, \citet{Battaglia:2012} found that halos with masses $M < 2\times 10^{14}\,M_{\odot}$ contribute half the tSZ power at $\ell = 3000$, with that fraction decreasing towards lower $\ell$.  Consequently, for $\ell < 3000$ (the range used to construct the $\kappa_{\rm CMB}$ maps from \citetalias{Omori:2017}), we expect our simulated map to capture more than 50\% of the tSZ power from halos on the sky.  We comment more on possible contributions to tSZ bias in the measured correlation functions from such low mass halos below.   There may also be tSZ signal on the sky that is not due to gas in massive halos, i.e. tSZ signal due to diffuse gas.  However, again, this contribution is expected to be subdominant to the contribution of the massive halos and would therefore not significantly change the estimated bias in $\kappa_{\rm CMB}$.

To assign tSZ profiles to the redMaPPer and \Planck clusters, we first estimate their masses, and then use a model to compute expected tSZ profiles given the estimated masses.  For the redMaPPer clusters, the masses are assigned using the mean mass-richness relation of \citet{Melchior:2017}.  For the \Planck clusters, the masses are assigned using the estimates constructed by \citet{Planck:clustercatalog} from the observed cluster tSZ signals.  In our fiducial analysis we set the hydrostatic bias parameter to $1-b=1$ when computing the masses of the \Planck clusters.  Given the mass estimates for the redMaPPer and \Planck clusters, we compute corresponding pressure profiles using the fits from \citet{Battaglia:2012}.  In particular, the thermal pressure profile is written as
\begin{equation}
\label{eq:battagliaprofile}
P_{\rm th}(x) = P_{200} P_0 (x/x_{\rm c})^{\gamma} \left[1 + (x/x_{\rm c})^{\alpha} \right]^{-\beta},
\end{equation}
where $x = r/R_{200c}$ and $R_{200c}$ is the radius from the cluster at which the enclosed mass is $M_{200c}$ and the  corresponding mean density is $3M_{200c}/(4\pi R_{200c}^3) = 200\rho_{\rm crit}(z)$.  The normalization, $P_{200}$ is given by
\begin{equation}
P_{200} = 200 \frac{GM_{200c}\rho_{\rm crit}(z) f_{\rm b}}{2 R_{200c}},
\end{equation}
where $f_{\rm b} = \Omega_{\rm b}/\Omega_{\rm m}$.  The parameters $P_0$, $x_{\rm c}$, $\alpha$, $\beta$, and $\gamma$ in Eq.~\ref{eq:battagliaprofile} are related to the cluster mass, $M_{200c}$, and redshift as described in \citet{Battaglia:2012}. The pressure profile is then converted to a Compton-$y$ profile by integrating along the line of sight,
\begin{equation}
y(\theta, M_{200c},z)= \frac{\sigma_{\rm T}}{m_{\rm e}c^{2}}\int dl \, P_{\rm e} (\sqrt{l^{2}+d_{\rm A}^{2}\theta^{2}},M_{200c},z),
\end{equation}
where $\sigma_{\rm T}$ is the Thomson cross-section, $m_{\rm e}$ is the electron mass, and the term in the integral is the electron pressure ($l$ is the line of sight distance, $d_{\rm A}$ is the angular diameter distance and $\theta$ is the angular separation relative to the cluster center). We assume that the electron pressure, $P_{\rm e}$, is given by $P_{\rm e} = 0.518  P_{\rm th}$.  This relation holds when the hydrogen and helium are fully ionized, and the helium mass fraction is $Y = 0.24$. 

Finally, the tSZ temperature signal at frequency $\nu$ is related to $y$ via
\begin{equation}
\frac{\Delta T(\nu)}{T_{\rm CMB}} = g\left(\frac{h\nu}{k_{\rm B} T_{\rm CMB}} \right) y,
\end{equation}
where $g(x) = x(e^x + 1)/(e^x-1)-4$ in the limit that the gas is non-relativistic \citep[e.g.][]{Carlstrom:2002}. 

In contrast to the \Planck and DES-detected clusters, for the SPT clusters we have a direct measurement of their tSZ profiles, and so use these measurements rather than modeling the profile through an estimate of the cluster masses.  \citet{Bleem:2015} performed fits to the observed $y$ profiles using the isothermal $\beta$ model \citep{Cavaliere:1976}, with $\beta = 1$:
\begin{equation}
\Delta T(\theta) = \Delta T_0 (1 + \theta/\theta_c)^{-1},
\end{equation}
where $\theta$ is the angular distance to the cluster and $\Delta T_0$ and $\theta_c$ are parameters of the fit.  For the SPT-detected clusters, we use these $\beta$-profile fits to estimate their contribution to the $y$ signal on the sky.  For any SPT-detected cluster that is also detected by \Planck or redMaPPer, we use the SPT measurement of its tSZ profile.

\begin{figure*}
\begin{center}
\begin{tabular}{cc}
\hspace{-0.5cm}
\includegraphics[width=0.5\textwidth]
{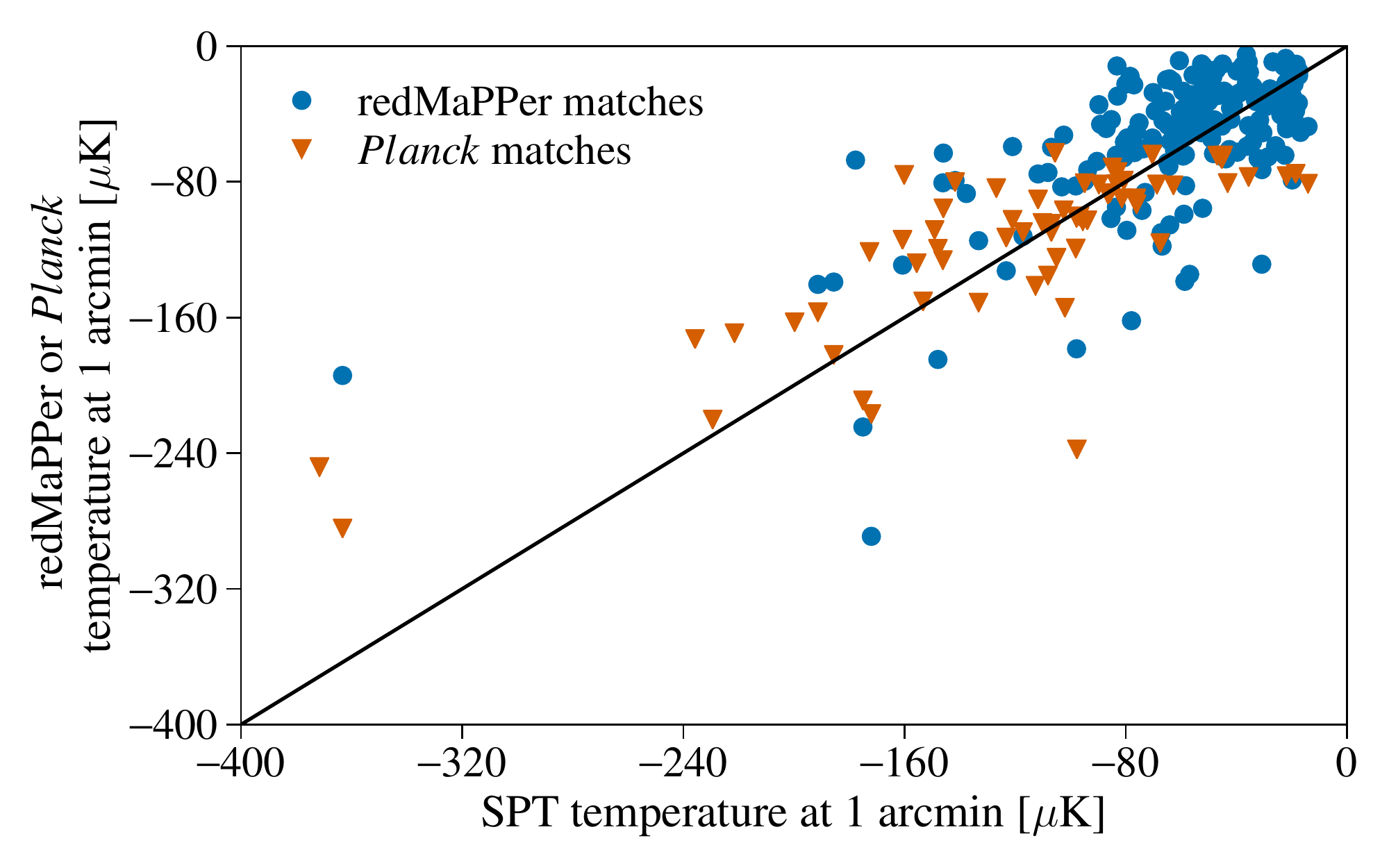}
\includegraphics[width=0.5\textwidth]
{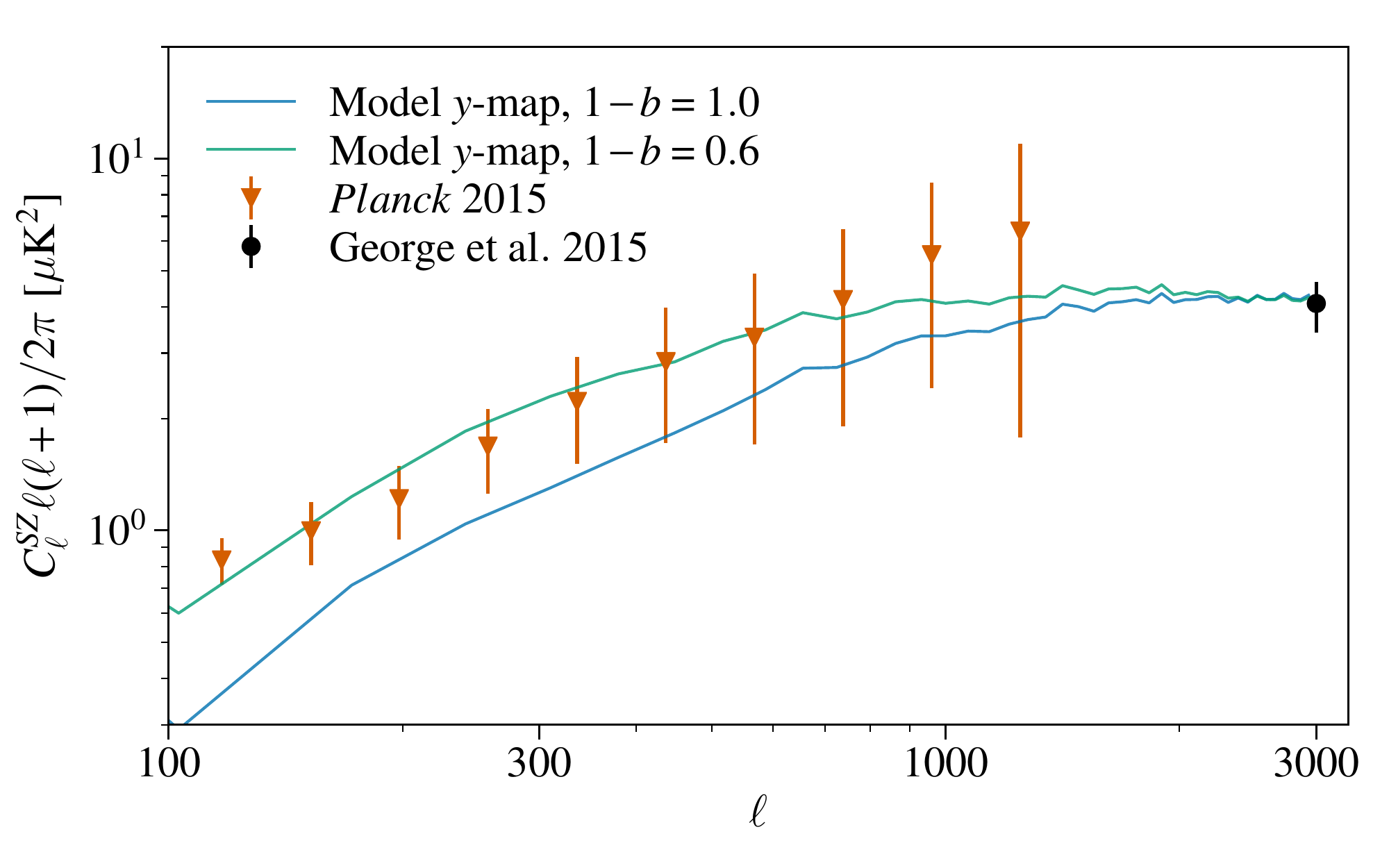}
\end{tabular}
\caption{The two panels show different tests of the simulated tSZ map used to estimate the effects of tSZ bias in the $\kappa_{\rm CMB}$ map of \citetalias{Omori:2017}.  The simulated map is generated by placing mock tSZ profiles at the locations of galaxy clusters detected by DES, SPT and \Planck.  (Left) Comparison of the amplitudes of the mock tSZ profiles of clusters detected in the different catalogs.  The $x$-axis represents the tSZ decrement at 150 GHz computed using the $\beta$-profile fits of \citet{Bleem:2015} to SPT-detected clusters, evaluated at one arcminute from the cluster center.  $y$-axis represents the same quantity computed for redMaPPer (blue circles) and \Planck-detected (red triangles) clusters using the \citet{Battaglia:2012} profile model described in the text.  The direct $y$-profile measurements from \citet{Bleem:2015} agree well with the estimated profiles for those clusters that appear in both the SPT catalog and the redMaPPer and \Planck catalogs.  (Right) Power spectrum of the simulated tSZ map compared to measurements from \citet{George:2015} and \citet{Planck:ymap}.  The two solid lines represent different assumptions about the masses of clusters that are detected by \Planck and not by SPT or DES.  As described in the text, the fiducial analysis assumes the $1-b=1.0$ model, but we find that the estimated bias is insensitive to this assumption.  This is not surprising, since the clusters that are only detected by \Planck live outside of the survey volume of DES, and the resultant bias is therefore largely uncorrelated with the DES galaxies and shears. Errorbars on the \Planck measurements include both statistical and foreground uncertainties \citep{Planck:ymap}.  The plot is restricted to modes with $100 < \ell < 3000$ since modes outside this range are not used in the $\kappa_{\rm CMB}$ reconstruction.   
}
\label{fig:tsz_power}
\end{center}
\end{figure*}

As a test of our simulated tSZ map, the left panel of Fig.~\ref{fig:tsz_power} shows a comparison of the estimated tSZ temperature profiles around the SPT, redMaPPer and \Planck clusters used to generate the tSZ map.  For those SPT-detected clusters that are also detected in the redMaPPer and \Planck catalogs, we plot the amplitude of the $\beta$-profile fits at one arcminute from the cluster center against the corresponding amplitudes of the estimated profiles from Eq.~\ref{eq:battagliaprofile}.  We choose to evaluate the profiles at one arcminute because this is roughly the beam scale of the SPT, so we do not expect the $\beta$-profiles to be well constrained below this scale.  The left panel of Fig.~\ref{fig:tsz_power} makes it clear that the estimated tSZ temperature profiles from Eq.~\ref{eq:battagliaprofile} agree well with the direct $\beta$-profile fits to the observed tSZ signals of the clusters.  This agreement is non-trivial: it provides a test of both of the profile model for the simulated tSZ map as well as the mass estimates for both the redMaPPer and \Planck clusters. 

As another check on the model $y$-profiles, we integrate the simulated profiles for the redMaPPer clusters out to $R_{500c}$ to obtain $Y_{500}$, and compare these values to the direct measurement of $Y_{500}$ around redMaPPer clusters from \citet{Saro:2017}.  \citet{Saro:2017} used a matched filter approach to estimate $Y_{500}$ for redMaPPer clusters detected in DES Science Verification data.  We find no evidence for a bias between the simulated and directly estimated $Y_{500}$ for richness $\lambda \gtrsim 60 $.  At richness $\lambda \lesssim 60$, we find that our model tends to yield higher $Y_{500}$ values, meaning that our model may be somewhat overestimating the effects of tSZ contamination.  Note that a similar discrepancy between the measured and predicted profiles was also found by \citet{Saro:2017}.  In that work, it was found that the measured $Y_{500}$ values for clusters with $\lambda < 80$ were smaller than predicted based on assumed scaling relations from \citet{Arnaud:2010}. 

As a further test of our simulated tSZ map, we compute the power spectrum of the map and compare the result to measurements of the $y$ power spectrum from \citet{George:2015} and \citet{Planck:ymap}.  This comparison is shown in the right panel of Fig.~\ref{fig:tsz_power}.  At $\ell = 3000$, our model yields a tSZ power spectrum that is in excellent agreement with that measured by \citet{George:2015}.  At $\ell \gtrsim 3000$, we expect the $y$ signal on the sky to receive significant contributions from low mass ($M \lesssim 2\times 10^{14}\,M_{\odot}$) and high-redshift halos ($z \gtrsim 0.6$) halos.  The fact that our simulated tSZ map does not include low-mass, high-redshift halos yet has power at $\ell = 3000$ that is as large as the \citet{George:2015} measurement suggests we may have somewhat overestimated the contribution to the $y$-signal from the redMaPPer clusters.  This explanation is consistent with the finding that our model predicts larger $Y_{500}$ values than measured by \citet{Saro:2017} for low richness clusters.  

For $\ell \lesssim 1000$, the tSZ power spectrum receives a significant contribution from clusters that are detected by \Planckc, and not by SPT or DES, i.e. high-mass, very low redshift clusters.  This can be seen from the fact that when we vary the hydrostatic mass bias parameter used to calculate masses for the \Planck clusters, the amplitude of the tSZ power spectrum at low $\ell$ changes significantly.  For our fiducial choice of $1-b=1.0$, we somewhat underpredict the tSZ power at low $\ell$; for $1-b=0.6$, we somewhat overpredict the tSZ power at low $\ell$, since this effectively assigns the \Planck clusters larger masses, and thus larger tSZ signals.  Although \citet{Planck:szcosmology} find evidence for $1-b=0.6$, this choice is not well motivated here since we are attempting to invert the SZ-derived masses to obtain an estimate of the corresponding SZ profiles.  Consequently, we keep $1-b = 1.0$ as the fiducial choice for the estimated tSZ map.  Note, though, that the amplitude of the inferred bias in $\galk$ and $\sheartk$ is almost completely insensitive to the value of $1-b$ that is assumed because the clusters that are only detected by \Planck are at very low redshift, and hence do not have strong correlations with DES galaxies or shears.  

\begin{figure*}
\begin{center}
\includegraphics[width=0.75\textwidth]{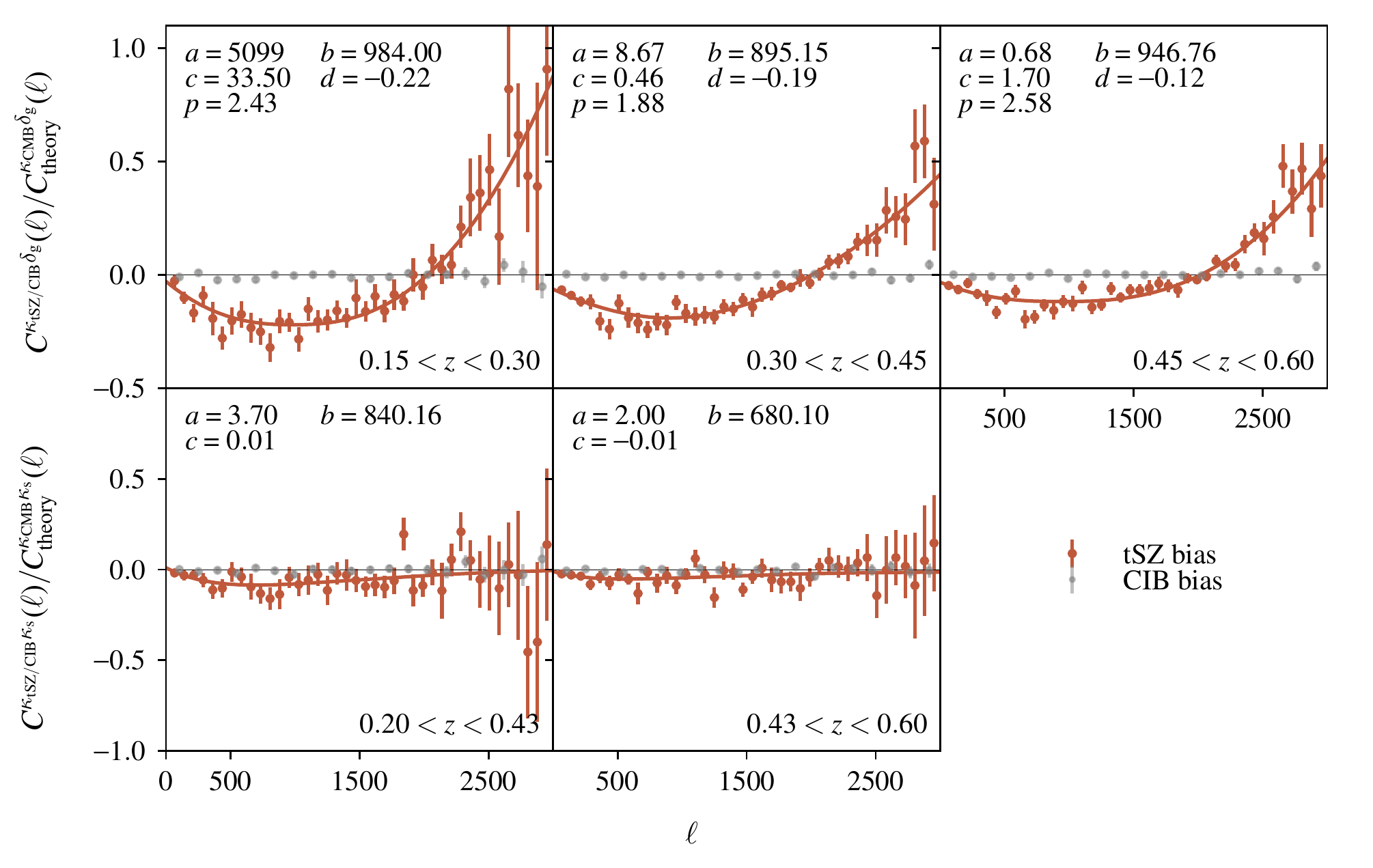}
\caption{The ratio of the $\delta_{\rm g}$ and $\gamma$ cross-correlations with the $\kappa_{\rm tSZ}$ map to the theoretical expectation for these correlations with the true $\kappa_{\rm CMB}$ map (prior to applying a Gaussian smoothing of FWHM=5.4').  These measurements form our estimate of the fractional bias in $\galk$ and $\sheartk$ due to tSZ contamination of the $\kappa_{\rm CMB}$ map from \citet{Omori:2017}.  Solid curves show model fits to Eqs. \ref{eq:tszbias_fitNK} and \ref{eq:tszbias_fitKG}, with the best-fit model parameters listed in each panel.  Grey points show equivalent quantities for the $\kappa_{\rm CIB}$ map. The  error bars shown are calculated using a spatial jackknife method. }
\label{fig:tszbiasfit}
\end{center}
\end{figure*}

\subsubsection{Masking clusters to reduce tSZ-induced bias}
\label{sec:masking}

Since galaxy clusters are sources of large tSZ signals, tSZ contamination of the $\kappa_{\rm CMB}$ maps can be reduced by masking these objects.   \citetalias{Omori:2017} masked clusters detected by SPT with signal-to-noise $\xi > 6$ when applying the quadratic lensing estimator to the SPT+\Planck CMB temperature maps.  Applying a more aggressive mask prior to the application of the quadratic estimator is problematic because a complicated mask will lead to difficulties with mode coupling.

In tests on the simulated $y$-map, we find that tSZ bias of the $\kappa_{\rm CMB}$ map can be further suppressed by masking additional clusters {\it after} the $\kappa_{\rm CMB}$ reconstruction.  This approach works because the application of the quadratic estimator with the filters defined in \citetalias{Omori:2017} to a localized tSZ source results in a somewhat-localized $\kappa_{tSZ}$ signal.  Masking clusters post-$\kappa$ reconstruction, then, can be used to reduce high-$\ell$ bias in the $\kappa_{\rm CMB}$ maps.  

Ultimately, the choice of clusters used for masking is set by the two competing desires to (a) reduce bias in $\galk$ and $\sheartk$ due to tSZ, while (b) ensuring that the bias induced by masking regions of high $\kappa_{\rm CMB}$ remains very small (see \S\ref{sec:maskingbias} for more discussion of this bias).  In tests on the simulated $y$-maps, we find that masking SPT-detected clusters with $\xi > 5$ and redMaPPer-detected clusters with $\lambda > 80$ post-$\kappa$ reconstruction can reduce the impact of tSZ bias while inducing an acceptable level of bias due to masking.  For all masked clusters, the mask radius employed is 5 arcminutes.  This choice of masking radius was found to significantly suppress the high $\ell$ bias from the tSZ  in tests on simulations, while simultaneously preserving most of the sky area.  The $\xi > 5$ masking threshold corresponds roughly to removing clusters with mass $M_{200m} \gtrsim 4\times 10^{14} M_{\odot}$ \citep{Bleem:2015}.  The $\lambda > 80$ threshold corresponds roughly to removing clusters with $M_{200m} \gtrsim 7\times 10^{14} M_{\odot}$ assuming the $\lambda$--$M$ relation from \citet{Melchior:2017}.  The fraction of sky area covered by the cluster mask is less than 1\%.

\subsubsection{Calculation of bias due to tSZ}

To estimate $\kappa_{\rm tSZ}$, we pass the simulated tSZ temperature map through the $\kappa_{\rm CMB}$ estimation pipeline of \citet{Omori:2017}.  We then correlate $\ktsz$ with the redMaGiC and shear catalogs described in \S\ref{sec:redmagic} and \S\ref{sec:shear} to estimate the biases in $\sheartk$ and $\galk$.  

We measure $\galktszell$ and $\kappasktszell$ in harmonic space using \texttt{PolSpice}.\footnote{\url{http://www2.iap.fr/users/hivon/software/PolSpice/}}  Fig.~\ref{fig:tszbiasfit} shows these bias functions relative to the theoretical expectation for $\galkell$ and $\kappaskell$ assuming the fiducial cosmological model shown in Table~\ref{tab:params}.  Although the exact values of the estimated biases are cosmology dependent, we are only attempting to determine the scales over which the tSZ bias is significant.  The change in these scales is negligible over the range of cosmological models allowed by the data.  The tSZ bias is well described by a multiplicative factor that is a smooth function of multipole, and which exhibits mild redshift dependence.  The bias in $\galkell$ is negative at scales of $\ell \lesssim 2000$, and positive for $\ell \gtrsim 2000$.  The amplitudes of these biases can be quite large, reaching a maximum of roughly 25\% for $\ell < 2000$, and even higher for $\ell > 2000$.  The tSZ bias in $\kappaskell$ does not exhibit as strong a peak at small scales as $\galkell$, but reaches similar levels of magnitude below $\ell \lesssim 2000$. 

Since the redMaPPer catalog is complete to only $z \sim 0.7$, we expect our estimate of the tSZ- induced bias in the last two redshift bins of $\galkell$ and $\kappaskell$ to be incomplete, since these bins receive contributions from structure at $z \gtrsim 0.7$. We therefore apply our bias measurements for the third-to-last redshift bin to the higher redshift bins.  We expect this approximation to be conservative, since the tSZ bias apparently decreases as a function of increasing redshift, as seen in Fig.~\ref{fig:tszbiasfit}.  This decrease is apparently physical, since the completeness of the redMaPPer and SPT catalogs does not evolve significantly over the redshift range $0.15 < z < 0.6$.

We fit the measured biases with smooth functions to make incorporation into our simulated analyses easier.  For the ratio of $C^{\ktsz\delg}(\ell)$/$C_{{\rm fid}}^{\kcmb\delg}(\ell)$, we find that the functions defined below provide a good fit:
\begin{equation}\label{eq:tszbias_fitNK}
y(\ell)=a(|(\ell-b)/c|)^{p}\times 10 ^{-8}+d,
\end{equation}
where $a$, $b$, $c$, $d$, and $p$ are free parameters for each redshift bin. Similarly, for $C^{\ktsz\kappa_s}(\ell)$/$C_{{\rm fid}}^{\kcmb\kappa_s}(\ell)$, we use a function of the form:
\begin{equation}\label{eq:tszbias_fitKG}
y(\ell)=-a\exp(-(\ell/b))^{1.2}\times10^{-4}+c.
\end{equation}
The results of these fits are shown as the solid curves in Fig.~\ref{fig:tszbiasfit}.  Given these parameterized fits, we can transform the biases measured in multipole space into biases in angular space (where $\galk$ and $\sheartk$ are measured). 

To assess how halos missing from the simulated tSZ maps could contribute to bias in the measured correlation functions, we repeat the bias estimates with different sets of halos masked.  We find that the contribution to the bias in the $\galk$ and $\sheartk$ correlation functions contributed by halos in the richness range $40 < \lambda < 80$ is larger than that from halos with $20 < \lambda < 40$ by roughly a factor of three.  Extrapolating this behavior to lower richness clusters suggests that massive halos with $20 \gtrsim \lambda \gtrsim 5$ do not contribute significantly to the bias.  Furthermore, we expect the tSZ contribution from halos with $M \lesssim {\rm few}\, \times 10^{13} M_{\odot}$ to be dominated by higher mass halos over all angular scales, given the beam size of SPT \citep[see, e.g.][]{Vikram:2017}.  These two arguments suggest that we have captured the majority of potential tSZ bias by using redMaPPer clusters with $\lambda > 20$ to generate the simulated tSZ map.

As seen in Fig.~\ref{fig:tszbiasfit}, the estimated biases due to tSZ leakage into the maps of $\kappa_{\rm CMB}$ are significant.  In \S\ref{sec:scale_cuts} we will assess the impact of these biases on the inferred cosmological constraints, and will choose scale cuts to mitigate their impact.

\subsection{Estimate of CIB bias}
\label{sec:cibbias}

We expect bias in the $\sheartk$ and $\galk$ correlation functions due to CIB bias to be small compared to the tSZ-induced bias.  Since the CIB is sourced predominantly from redshifts $z \sim 2$, it is not expected to correlate strongly with the galaxy or shear samples used in this work.  We now attempt to confirm this expectation.

We estimate the effects of CIB contamination of the $\kappa_{\rm CMB}$ maps on $\galk$ and $\sheartk$ using a procedure similar to that used to estimate the tSZ bias.  However, rather than generating a simulated CIB map, we instead rely on \Planck observations.  To this end, we use the \Planck GNILC 545 GHz CIB map \citep{Planck:CIB} as a proxy for the true CIB emission on the sky.  We first calculate the $\ell$-dependent cross-correlation between the combined SPT+\Planck map and the \Planck GNILC $545$ GHz maps; this correlation provides an estimate of the amount of CIB contamination in the SPT+\Planck map. The GNILC $545$ GHz map is then convolved with the $\ell$ dependent scaling function: 
\begin{equation}
\eta(\ell)=\frac{C_{\ell}^{{\rm GNILC}\times {\rm SP}}}{C_{\ell}^{{\rm GNILC}\times{\rm GNILC}}},
\end{equation}
where SP refers to the SPT+\Planck map.  The result is a map of the estimated CIB leakage into the SPT+\Planck temperature map.  

Next, the quadratic estimator is applied to the estimated CIB leakage map to produce $\kappa_{\rm CIB}$, an estimate of the leakage of CIB into the $\kappa_{\rm CMB}$ map.  As with $\kappa_{\rm tSZ}$, we cross-correlate $\kappa_{\rm CIB}$ with the true DES galaxy and shear catalogs to form estimates of the bias in $\sheartk$ and $\galk$ due to CIB leakage.  These cross-correlations are shown in Fig.~\ref{fig:tszbiasfit}.  From the figure, it is apparent our estimate of the CIB bias is consistent with there being no bias, and we will henceforth ignore CIB as a potential source of contamination in our analysis.

\subsection{Biases due to masking clusters}
\label{sec:maskingbias}

As mentioned in \S\ref{sec:masking}, massive galaxy clusters are masked to reduce contamination of  $\kappa_{\rm CMB}$ by tSZ leakage.  However, clusters are also associated with regions of high $\kappa_{\rm CMB}$.  Consequently, by masking these objects, we expect to reduce the amplitude of $\sheartk$ and $\galk$ somewhat, which could result in a bias to parameter constraints.  Furthermore, masking regions of high signal can also change the behavior of estimators for that signal.  For these reasons, we have not attempted to reduce the tSZ bias any further with more extreme masking.  We prefer instead to ensure that the masking bias remains negligible, as we will show below.  Note that the total masked area is quite small because there are relatively few clusters on the sky.  Less than 1\% of the pre-masking survey area is removed by the cluster mask, which masks 437 clusters.  

To characterize masking-induced bias, we generate a simulated $\kappa_{\rm CMB}$ map that consists only of mock cluster $\kappa_{\rm CMB}$ profiles at the locations of the masked clusters in the data; we refer to this map as $\kappa_{\rm sim}$.  Each cluster is modeled with an Navarro-Frenk-White (NFW) profile \citep{Navarro1996}.  Taking a somewhat simplistic approach, we assign each simulated cluster a mass of $10^{15}\,M_{\odot}$, which we expect to overestimate the effects of the masking, since most of the masked clusters will have masses less than this.   The simulated $\kappa_{\rm sim}$ map is then correlated with the true galaxy and shear catalogs to estimate $\galksim$ and $\sheartksim$.\footnote{In practice, masked pixels are excluded from the analysis when computing correlation functions.  Our estimate of the masking bias, however, corresponds instead to setting these pixels to zero. Given the small angular size of the masked clusters, the difference between these two approaches should be small.  If anything, we {\it overestimate} the effects of masking by computing the bias in this manner.}  These two correlation functions effectively represent the parts of $\galk$ and $\sheartk$ that we have "missed" by masking the massive galaxy clusters. We find that the ratios of $\sheartksim$ and $\galksim$ to the true correlation functions are approximately constant with angular scale, and have an average amplitude of approximately 1\%.  A 1\% bias is significantly below the bias induced by e.g. tSZ, and we will therefore ignore it in the subsequent analysis.  The level of bias induced by masking is schematically illustrated by the dashed line in Fig.~\ref{fig:sys_physical_dist}.

\section{Choice of Angular Scale cuts}
\label{sec:scale_cuts} 
 
\begin{figure*}
\begin{center}
\includegraphics[width=1.0\textwidth]{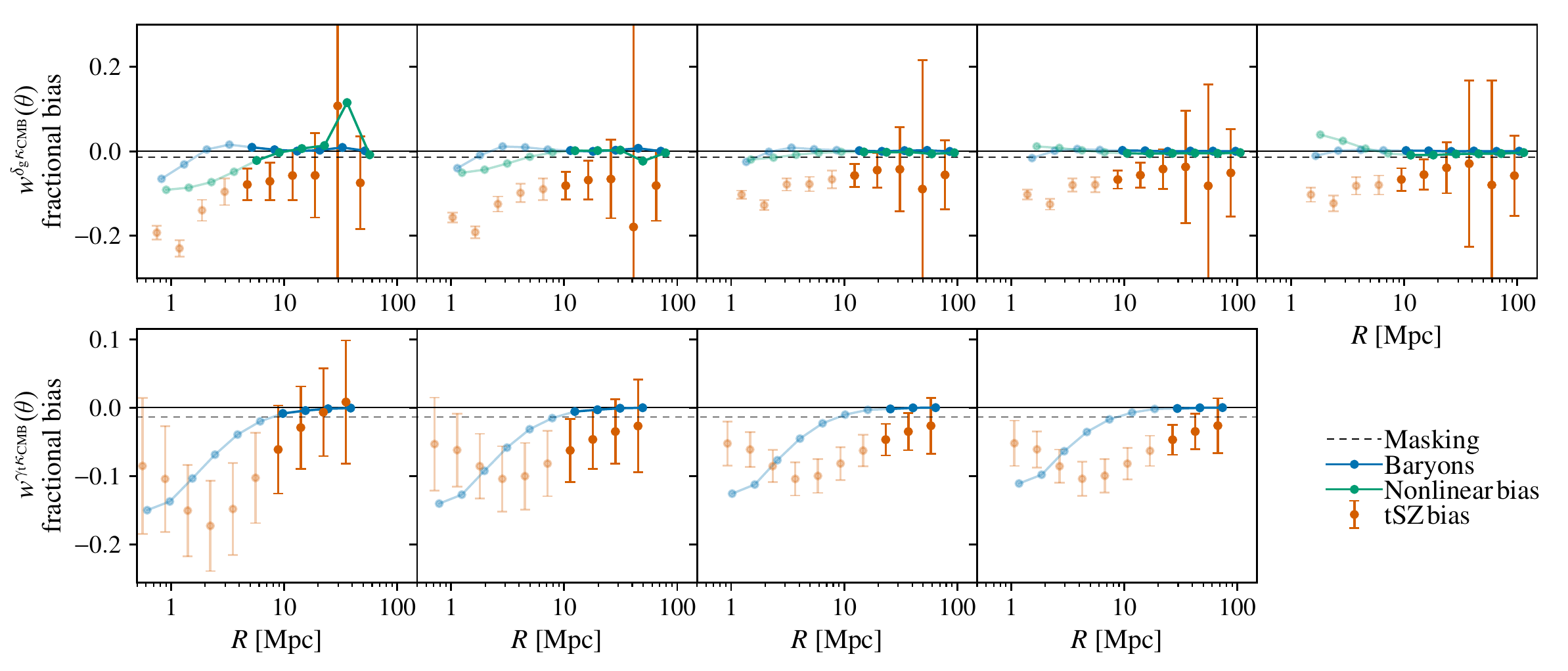}
\caption{Biases in $\sheartk$ and $\galk$ relative to the error bars as a function of physical separation.  Faded points are excluded by scale cuts. Errorbars correspond to 10\% of the square root of the diagonal of the covariance matrix described in \S\ref{sec:covariance}; for ease of visualization, we only plot errorbars on the tSZ-biased points.  The dashed line labeled `Masking' refers to the roughly 1\% bias induced by masking galaxy clusters described in \S\ref{sec:maskingbias}.}
\label{fig:sys_physical_dist}
\end{center}
\end{figure*}

\begin{figure*}
\begin{center}
\begin{tabular}{cc}
\hspace{-0.5cm}\includegraphics[width=0.49\textwidth]{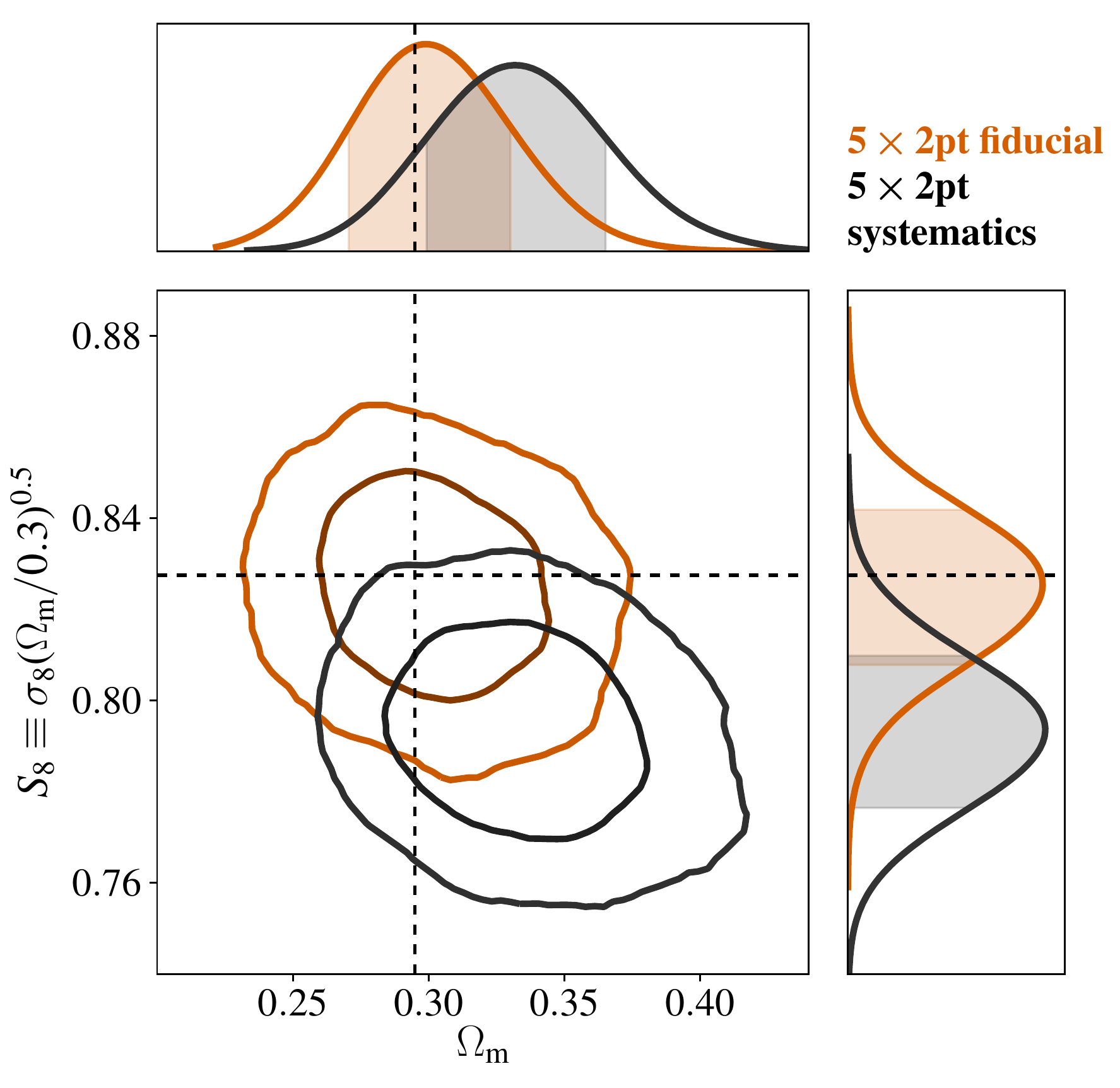}
\includegraphics[width=0.49\textwidth]{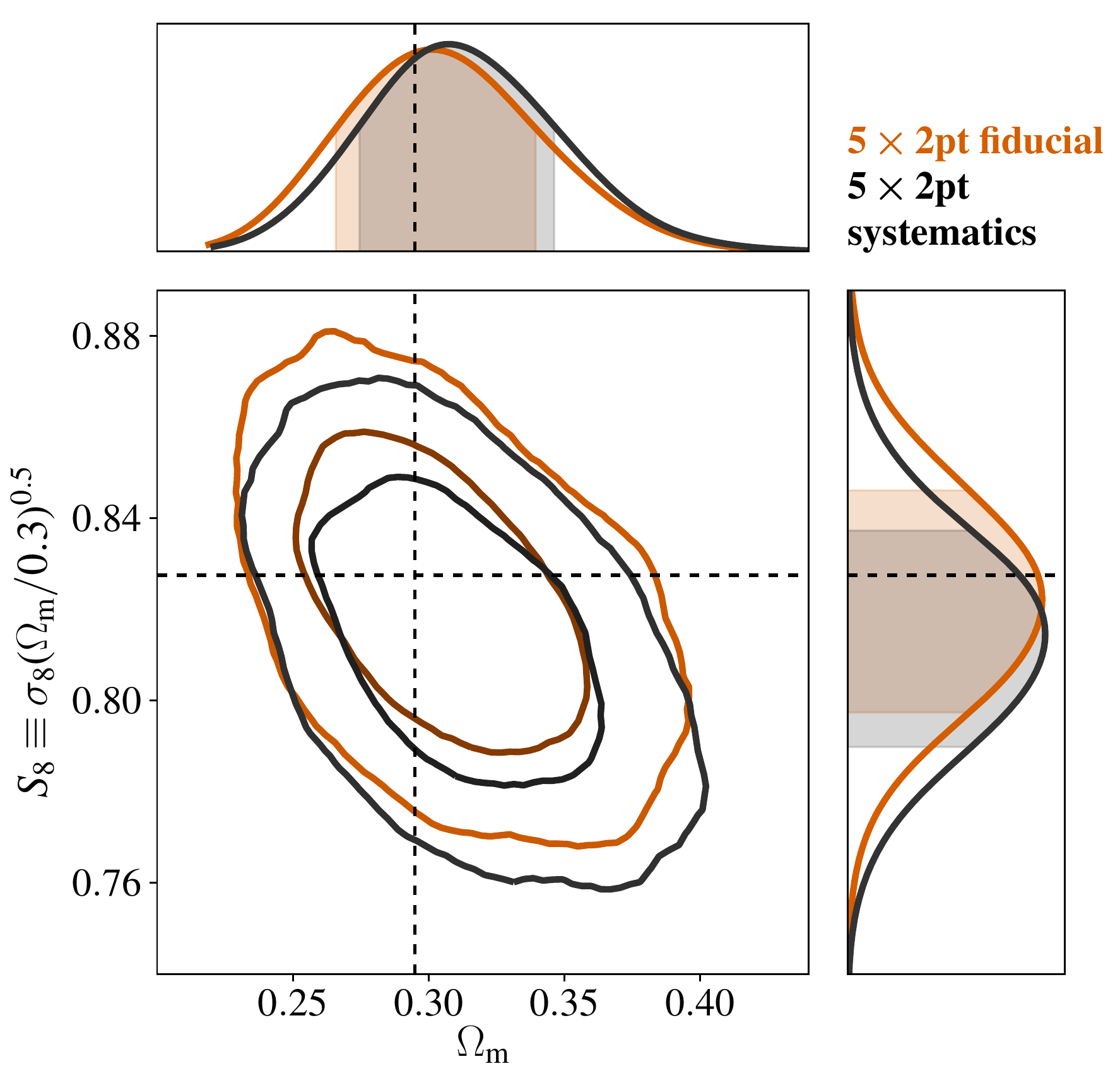}
\end{tabular}
\caption{Effects on cosmological constraints of unmodeled contributions to the simulated data vector before (left) and after (right) the application of angular scale cuts on $\galk$ and $\sheartk$.  `Fiducial' refers to the data vector generated using the baseline model described in \S\ref{sec:modeling}; `Systematics' refers to the simulated data vector that includes prescriptions for tSZ bias in the $\kappa_{\rm CMB}$ map, nonlinear galaxy bias, and the OWLS AGN model for baryons.  The scale cuts applied to the \3x2pt subset of observables are kept fixed throughout to those of \citet{DESy1:2017}.}
\label{fig:scale_cuts}
\end{center}
\end{figure*}

When modeling the \5x2pt data vector, we neglect nonlinear galaxy bias, the impact of baryons on the matter power spectrum, and the presence of tSZ contamination in the $\kappa_{\rm CMB}$ maps.  To prevent these unmodeled effects from causing biases in our cosmological constraints, we restrict our analysis to scales over which their impact is small.  In general, these effects become significant at small scales, so this restriction is tantamount to removing small scales from the analysis.   

We follow the same basic approach for determining the scale cuts as in \citetalias{Krause:2017}: we introduce estimates of the unmodeled effects into a simulated data vector generated at the fiducial parameter values from Table~\ref{tab:params}, and analyze this data vector with varying scale cuts to determine how the parameter constraints are impacted.  If the impact of these effects is sufficiently small, we consider our choice of scale cuts sufficient.  Our heuristic threshold for an acceptable bias is that the bias on any parameter should not be larger than 50\% of the statistical uncertainty on that parameter.  The resultant scale cuts reduce the bias in the cosmological constraints to acceptable levels, but at the cost of increasing our statistical errorbars.  Future work will be devoted to improving modeling of nonlinear bias, baryons and tSZ bias in order to exploit the additional statistical power in the data. 

For the \3x2pt subset of observables, we adopt the same scale cuts as in \citetalias{Krause:2017}.  In principle, the improved signal-to-noise from including $\galk$ and $\sheartk$ in the analysis could necessitate more conservative scale cuts for the \3x2pt subset.  However, we find below that this is not necessary.  

To determine scale cuts for the $\galk$ and $\sheartk$ correlation functions, we consider the impact of three systematics that are expected to dominate: tSZ bias in the $\kappa_{\rm CMB}$ maps, nonlinear galaxy bias, and the effects of baryons.  Of these, we find that tSZ bias in $\kappa_{\rm CMB}$ is generally dominant.  We introduce these effects into the simulated data vectors in the following fashion:
\begin{itemize}
\item \textit{tSZ bias:} tSZ bias is introduced into the simulated data vector using the harmonic space fits described in \S\ref{sec:kappa_systematics} and shown in Fig.~\ref{fig:tszbiasfit}.  
\vspace{0.2cm}
\item \textit{Nonlinear galaxy bias:} following \citetalias{Krause:2017}, we compute the corrections to $\galk$ resulting from the next to leading order bias correction, $b_2$, and tidal bias term, $b_{\rm s}$ \citep{McDonald:2009,Baldauf:2012}.  These terms are computed using FAST-PT \citep{McEwen:2016}. 
\vspace{0.2cm}
\item \textit{Baryons:} following \citetalias{Krause:2017}, we introduce baryonic effects into the simulated data vector using results from the OWLS simulations \citep{Schaye:2010}.  In particular, we use the OWLS AGN model, which is expected to provide an upper limit to the effects of baryons on the matter power spectrum. The modifications to the power spectrum due to baryons are propagated into the mock data vectors using Eqs.~\ref{eq:CGK} and \ref{eq:CNK}.
\end{itemize}

A potential source of systematic bias considered by \citetalias{Krause:2017} was the impact of a one-halo term on $\galsheart$.  Since $\galsheart$ mixes power from small scales into large scales, the one-halo term can impact $\galsheart$ at scales significantly beyond the halo virial radius.  In contrast, $\galk$ at a projected distance $R$ from halos depends only on the matter power at scales larger than $R$.  Since we exclude small scales of $\galk$ anyway, it is safe to ignore the effects of the one-halo term on $\galk$ in this analysis. 

Fig.~\ref{fig:sys_physical_dist} shows the fractional changes in $\galk$ and $\sheartk$ induced by tSZ bias, nonlinear galaxy bias, and the OWLS baryon model.  For $\galk$, we plot the fractional change as a function of the projected physical separation evaluated at the mean redshift of the tracer galaxies.  For $\sheartk$, we plot the fractional change as a function of the projected physical separation evaluated at the peak of the lensing kernel of the source galaxies. The errorbars plotted in Fig.~\ref{fig:sys_physical_dist} are intended to allow comparison between the bias and the statistical uncertainties of the measurements; for better visualization, the errorbars correspond to only 10\% of the square root of the diagonal of the covariance matrix.  For each angular bin, the bias is not highly significant, but the combined effect from all bins is significant, as we show below. 

Fig.~\ref{fig:sys_physical_dist} also makes it clear that over most scales, tSZ contamination is the most significant source of bias in our analysis.  Note that baryons have a fairly small impact on $\galk$, and the nonlinear bias does not impact $\sheartk$ at all since this correlation function does not involve biased tracers of the mass.  Below scales of about 3 Mpc, bias due to the impact of baryons begins to dominate over the tSZ-induced bias in $\sheartk$.  Clearly, though, removing tSZ bias from the $\kappa_{\rm CMB}$ maps would allow us to push the analysis to significantly smaller scales.

Our scale cut choice is also illustrated in Fig.~\ref{fig:sys_physical_dist}.
The faded points in the figure illustrate the scales that are removed from the analysis by the scale cuts.  We exclude angular scales below $(15',25',25',15',15')$ for the five redshift bins of $\galk$, and below $(40',40',60',60')$ for the four redshift bins of $\sheartk$.  For $\galk$, the cuts correspond roughly to restricting to scales $R > 8$ Mpc, and somewhat smaller for the lowest redshift bin.  

We define the $\Delta \chi^2$ between the biased and unbiased data vectors as 
\begin{eqnarray}
\Delta \chi^2 = \left(\mathbf{d}_{\rm bias} - \mathbf{d}_{\rm fid} \right)^T \mathbf{C}^{-1} \left( \mathbf{d}_{\rm bias} - \mathbf{d}_{\rm fid} \right),
\end{eqnarray}
where $\mathbf{d}_{\rm bias}$ and $\mathbf{d}_{\rm fid}$ are the data vectors with and without the unmodelled effects, respectively.  Including all three unmodeled effects simultaneously, before the application of scale cuts, we find that for the combination of $\galk$ and $\sheartk$, $\Delta \chi^2 = 10.2$ (with $\nu = 90$ degrees of freedom).  After the scale cuts are imposed, $\Delta \chi^2$ for the $\galk$ and $\sheartk$ combination is reduced to only 0.26 (with $\nu = 43$ degrees of freedom).  We compute the effect of the residual $\Delta \chi^2$ on the parameter constraints below.

Using the MCMC methods described in \S\ref{sec:likelihoodanalysis}, we compute the posteriors on the full set of model parameters with and without the unmodeled sources of bias, and with and without the imposition of the scale cuts.  These results are shown in Fig.~\ref{fig:scale_cuts}.  For ease of visualization, we show the shifts in the posteriors only in the space of $\Omega_{\rm m}$ and  $S_8$.  These two cosmological parameters are tightly constrained by the \3x2pt and \5x2pt analysis, and so are particularly useful for assessing the effectiveness of our scale cut choices.  The left panel of Fig.~\ref{fig:scale_cuts} shows the  constraints on $\Omega_{\rm m}$ and $S_8$ obtained when analyzing the simulated data vectors with and without the unmodeled effects when all scales are included in the analysis of $\galk$ and $\sheartk$ (but imposing the fiducial scale cuts on the \3x2pt subset of the data vector).  In this case, the bias induced by the unmodeled effects is unacceptably large, significantly greater than the statistical uncertainties.  The right panel of Fig.~\ref{fig:scale_cuts} shows the cosmological constraints when small scales are excluded as described above.  In this case, the bias is significantly reduced at the cost of larger error bars.  We find that the shift in the 68\% confidence interval for $S_8$ due to the unmodeled effects is roughly 38\% of the statistical uncertainty on $S_8$, which we deem acceptably small.  The shift in the mean $\Omega_{\rm m}$ is 23\% of the statistical uncertainty on $\Omega_{\rm m}$.  We also note that with the scale cuts imposed, $\Omega_{\rm m}$ appears to be degenerate with $S_{8}$, while they are much less degenerate without the scale cuts. This implies that the additional small-scale power in the $\sheartk$ and $\galk$ measurements helps to break this degeneracy. Note that the residual bias exhibited in the right panel of Fig.~\ref{fig:scale_cuts} is partially due to the effects of nonlinear galaxy bias and baryons on the \3x2pt combination of observables.  The total $\Delta \chi^2$ between the biased and fiducial \5x2pt data vectors is 0.81.  Of this, 0.45 is contributed by $\galk$ and $\sheartk$.  One could in principle make the \3x2pt scale cuts more conservative in order to relax the scale cuts on $\galk$ and $\sheartk$ somewhat.  However, we have not taken this approach in order to maintain consistency with the analysis of \citet{DESy1:2017}.

We note that our choice of scale cuts removes a significant fraction of the signal-to-noise in $\sheartk$ and $\galk$, resulting in significantly degraded cosmological constraints from these two correlation functions.  However, given that we use the $\kappa_{\rm CMB}$ maps from \citet{Omori:2017}, this choice seems unavoidable.  For future work, reducing tSZ leakage into the $\kappa_{\rm CMB}$ maps is a high priority.  Alternatively, it may be possible to model the effects of tSZ bias in the analysis.

\section{Results of simulated analyses}
\label{sec:forecasts}

Having described our model for the \5x2pt combination of observables and our choice of angular scale cuts, we now present the results of simulated likelihood analyses.  For this purpose, we use the simulated data vector described in \S\ref{sec:scale_cuts}.  The simulated data vector is generated without noise so that --- by definition --- the maximum likelihood point occurs at the true parameter values.  

\begin{figure}
\includegraphics[width=0.49\textwidth]{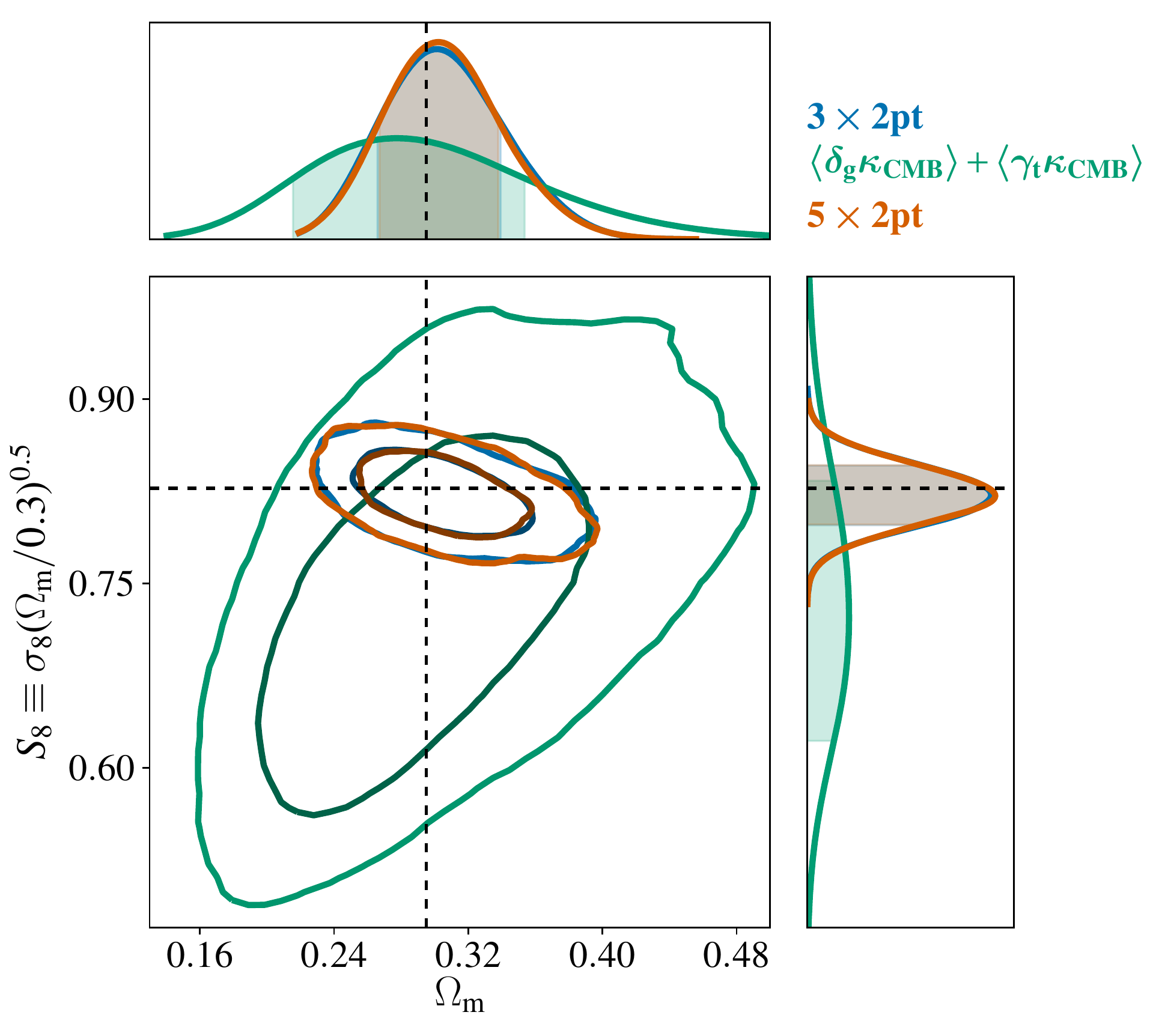}
\caption{Constraints on $\om$ and $S_8$ for \3x2pt (red), \5x2pt (blue), and the two 2pt function that cross-correlation with the CMB lensing map, $\galk$ and $\sheartk$ (green). The dashed black line shows the fiducial values of $\om$ and $S_{8}$.}
\label{fig:fid_constraints}
\end{figure}

\begin{figure}
\includegraphics[width=0.49\textwidth]{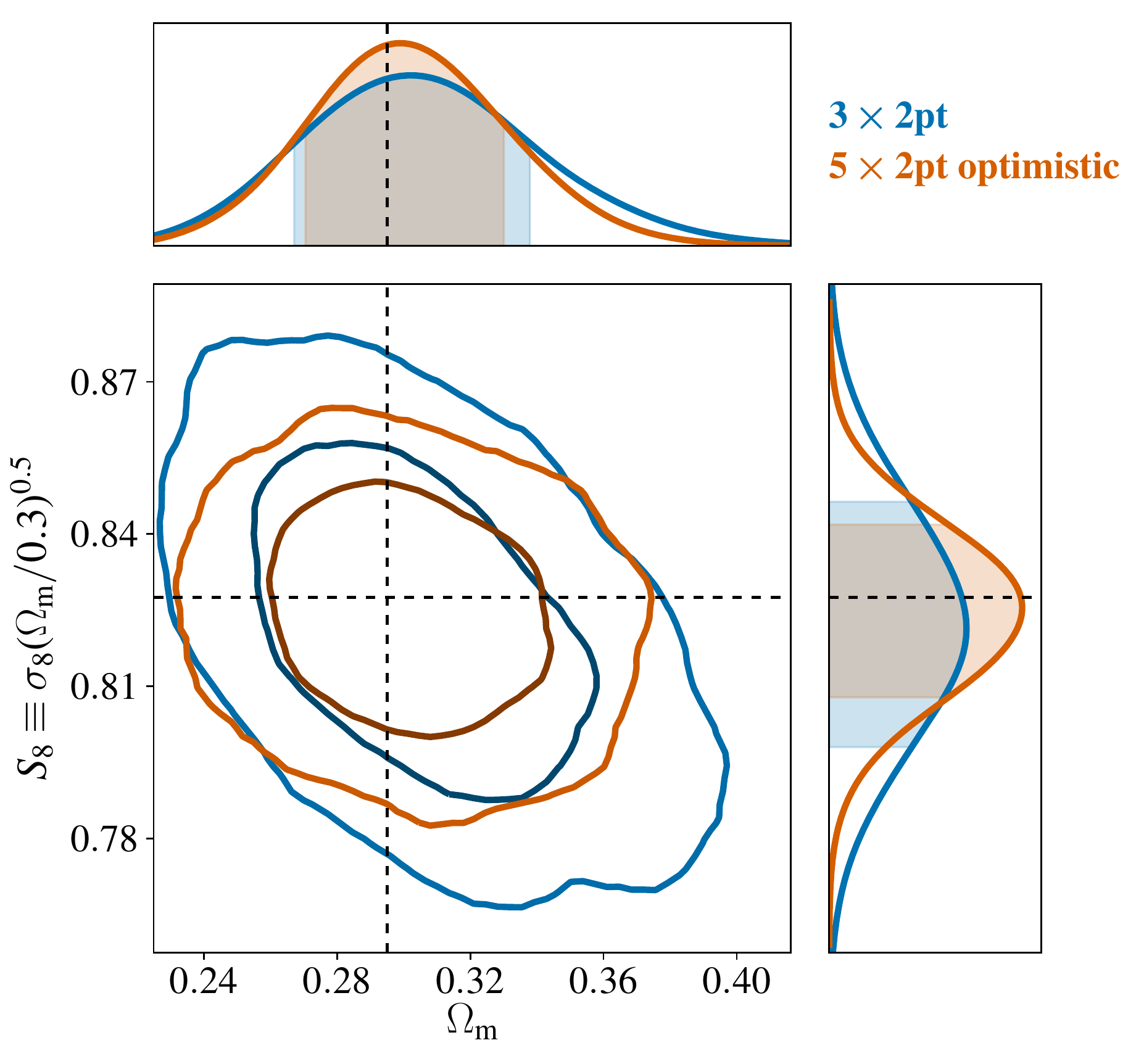}
\caption{Constraints on $\om$ and $S_8$ for \3x2pt (red) and \5x2pt (blue) when no scale cuts are applied to the analysis. The dashed black line shows the fiducial values of $\om$ and $S_{8}$. }
\label{fig:noscalecuts}
\end{figure}

\subsection{Fiducial results}
\label{sec:fid_results}

We first present projected constraints on cosmological parameters generated from our analysis of a simulated \5x2pt data vector assuming the fiducial choice of angular scale cuts described in \S\ref{sec:scale_cuts}.  Fig.~\ref{fig:fid_constraints} shows the constraints on $\om$ and $S_8$ generated from our fiducial analysis under the $\Lambda$CDM model.  Also shown in Fig.~\ref{fig:fid_constraints} is the constraint coming from the joint analysis of $\galk$ and $\sheartk$ alone.  Given the current errorbars, the constraining power of $\galk$ and $\sheartk$ is significantly weaker than that of the \3x2pt combination.  This is not too surprising given the low signal-to-noise of the $\galk$ and $\sheartk$ correlation functions after the imposition of scale cuts: the combined signal-to-noise from these observables is roughly 8.8.  The signal-to-noise of the \3x2pt combination after imposing scale cuts, on the other hand, is approximately 41.  Consequently, extending \3x2pt to \5x2pt does not have a dramatic impact in terms of tightening cosmological constraints.  Interestingly, though, the degeneracy direction of the combined $\galk$ and $\sheartk$ constraint in the $\om$--$S_8$ parameter space is very complementary to that of the \3x2pt analysis.

Ignoring the effects of tSZ, nonlinear galaxy bias, and baryons, the projected signal-to-noise of the \5x2pt analysis including all angular bins is 20.  After the fiducial scale cuts are imposed, the signal-to-noise is reduced to 8.8.  An interesting question to ask, then, is how well could the \5x2pt combination constrain cosmology if all of the original signal to noise could be exploited?  Fig.~\ref{fig:noscalecuts} shows the cosmological constraints from the \5x2pt analysis on $S_8$ and $\om$ when no scale cuts are imposed on $\galk$ and $\sheartk$.  In this case, the \5x2pt analysis significantly shrinks the constraint contour.  We note that this figure is meant simply to illustrate the potential signal-to-noise of the cross-correlations between DES Y1 data and the $\kappa_{\rm CMB}$ maps.  The result is overly optimistic because it ignores other sources of model bias (i.e. baryons, nonlinear galaxy bias, etc.).  As shown in Fig.~\ref{fig:sys_physical_dist}, other sources of model bias can become significant at small scales.  All results presented below will use the fiducial choice of scale cuts described in \S\ref{sec:scale_cuts}.

\subsection{Self-calibration of systematics parameters}
\label{sec:nuisance_constraints}

In addition to the cosmological parameters, there are many  nuisance parameters varied in this analysis, including $m_i$, $\Delta z_{\rm s}$, the galaxy bias, and intrinsic alignment parameters.  One of the main advantages of joint two-point function analyses is that the resultant cosmological constraints are quite robust to such nuisance parameters \citep[e.g.][]{Hu:2004}.  This is not true for the analysis of single 2pt functions. For example, fits to $\galsheart$ alone lead to complete degeneracy between galaxy bias and $A_{\rm s}$, while fits to $\xi_{+/-}(\theta)$  lead to a complete degeneracy between $m$ and $A_{\rm s}$.  Many of these degeneracies are broken by the \3x2pt combination of observables, since there is no nuisance parameter that affects $\wtheta$, $\galsheart$, and $\xi_{+/-}(\theta)$ in the same way.  For instance, $\galsheart$ scales with the shear calibration bias as $(1+m)$,  $\xi_{+/-}(\theta)$ scales with $(1+m)^2$, but $\wtheta$ is independent of $(1+m)$.

\begin{figure}
\includegraphics[width=\linewidth]{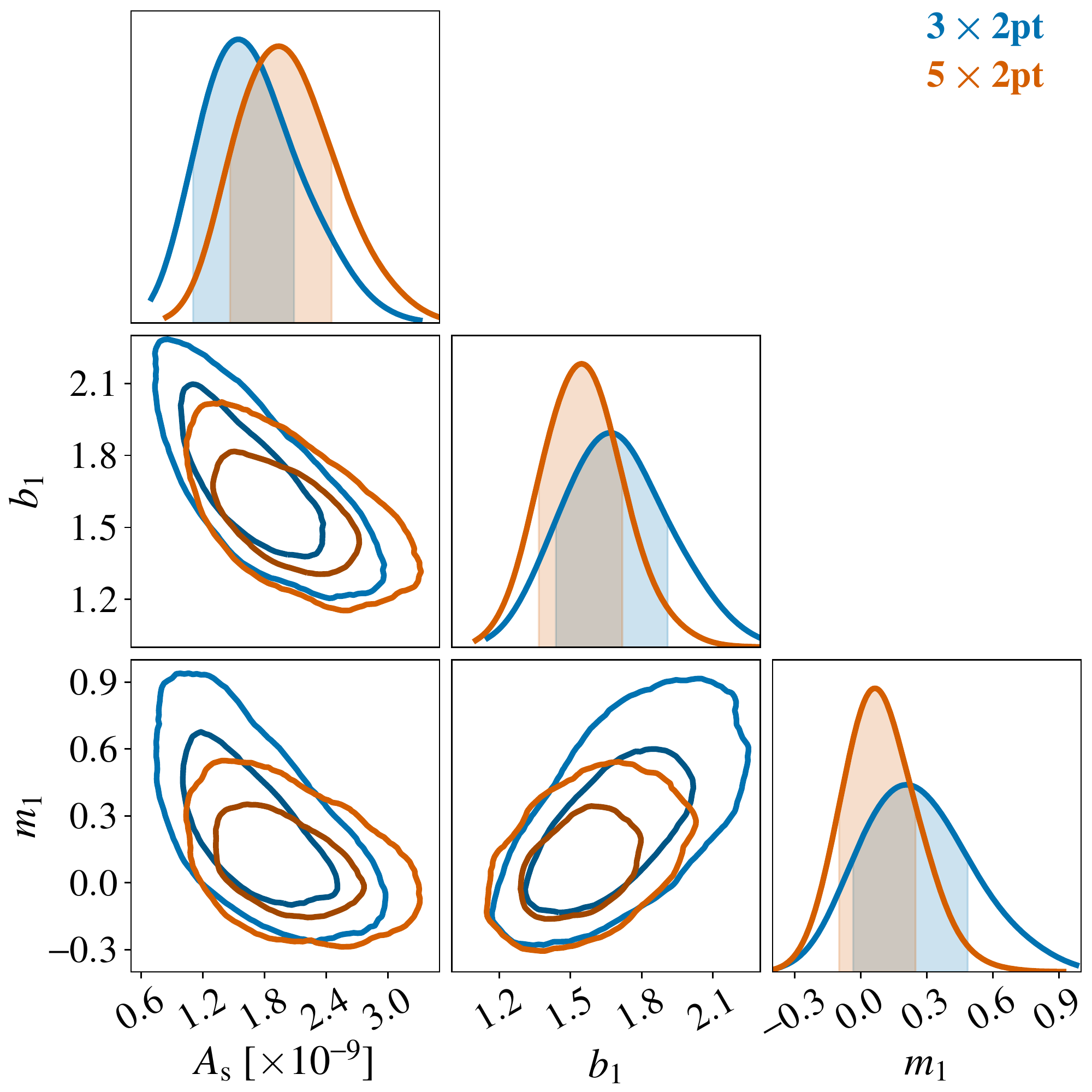}
\caption{Illustration of the degeneracy between $A_s$, galaxy bias ($b_1$) and shear bias ($m_1$) in the \3x2pt and \5x2pt analyses.  For this figure, we have placed non-informative priors on the shear calibration parameters.}
\label{fig:bm_degeneracy}
\end{figure}

However, even the \3x2pt analysis of \citet{DESy1:2017} is not completely immune to degeneracies between nuisance parameters and cosmological parameters.  In particular, the cosmological constraints of the \3x2pt analysis are degraded by a three-parameter degeneracy between galaxy bias, shear calibration, and $A_{\rm s}$.  Consider the effect of increasing the galaxy bias, $b$, by some factor $\alpha > 1$ such that $b \rightarrow \alpha b$.  In that case, the amplitude of $\galsheart$ will increase by $\alpha$ and $\wtheta$ will increase by $\alpha^2$, while $\xi_{+/-}(\theta)$ remains unchanged.  These changes can be compensated partly by decreasing $A_{\rm s}$ by $\alpha^2$, which will result in $\xi_{+/-}(\theta)$ decreasing by $\alpha^2$, $\galsheart$ being reduced by $\alpha$ relative to its original value, and $w(\theta)$ returning to its original value.  Finally, if shear calibration, $m$, is increased such that $(1+m) \rightarrow \alpha(1+m)$, then $\galsheart$ and $w(\theta)$ will return to their original values.  The net result is a counterintuitive {\it positive} correlation between $m$ and galaxy bias.  This degeneracy is illustrated for a single redshift bin with the blue contours in Fig.~\ref{fig:bm_degeneracy}.  Since the fiducial priors on $m$ significantly restrict its allowed range, it is hard to see the degeneracy between $m$ and other parameters when these priors are imposed.  Consequently, when generating Fig.~\ref{fig:bm_degeneracy} we have replaced the fiducial $m$ prior with one that is flat over the range $m \in [-1, 1]$.

\begin{figure}
\centering
\includegraphics[width=\linewidth]{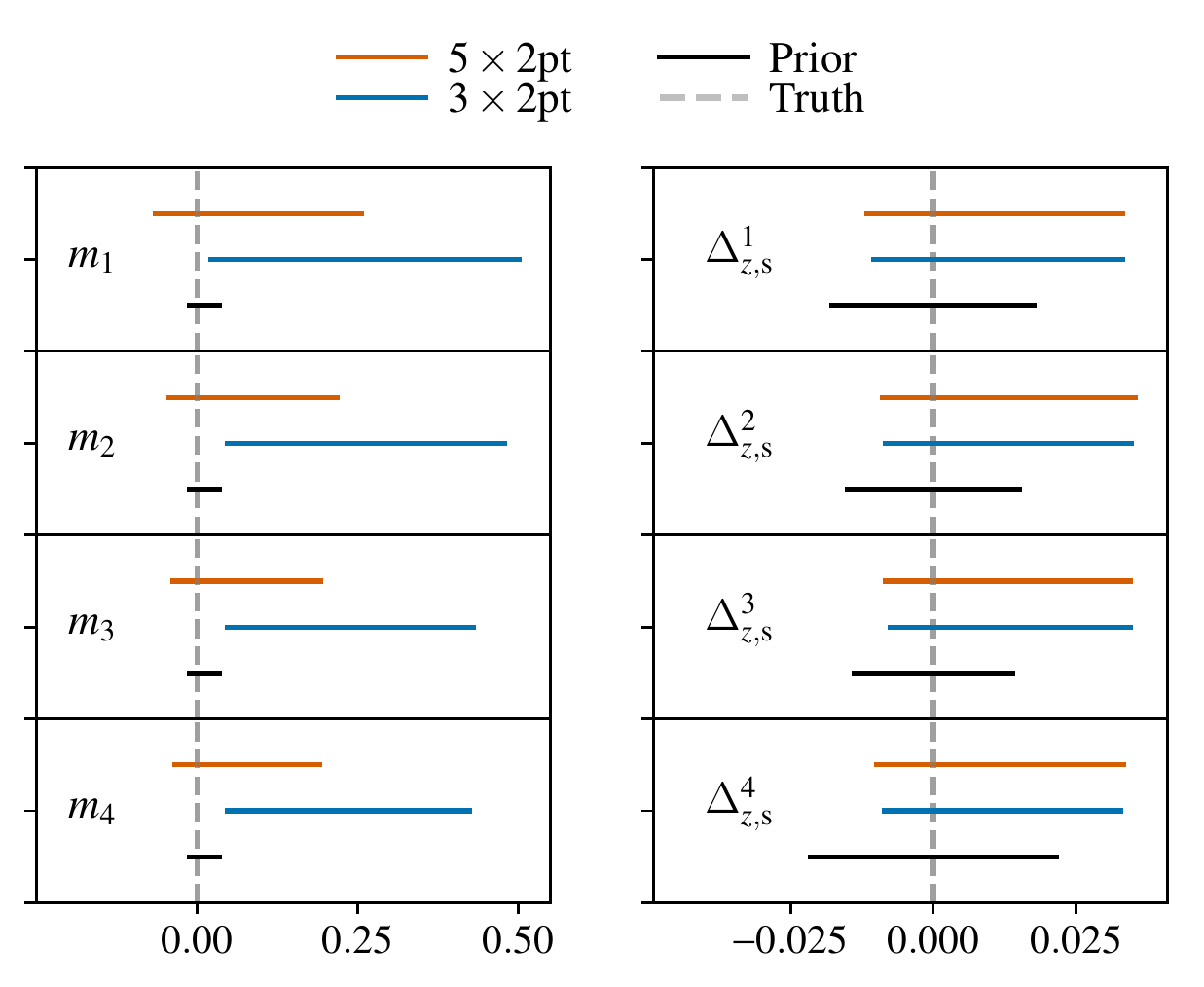}
\caption{Recovered constraints (68\% confidence interval) on multiplicative shear bias (left) and photometric redshift bias (right) for the \5x2pt analysis (orange bars) and \3x2pt analysis (blue bars) when the priors on these parameters are completely non-informative.  Black bars show the priors imposed on the $m_i$ in the fiducial analysis.}
\label{fig:mzsconstraint}
\end{figure}

As a result of the above degeneracy in the \3x2pt analysis, it is useful to impose informative priors on the multiplicative bias parameters and the redshift bias parameters.  For the \citet{DESy1:2017} analysis, the priors on multiplicative shear bias for the \textsc{MetaCalibration} catalog are derived using a variety of tests described in \citet{Zuntz17}.  In the case of redshift biases, priors on the source redshift biases are derived using comparisons to data from the COSMOS \citep{Laigle:2016} field in \citet{Hoyle2017} and angular clustering in \citet{Davis:2017} and \citet{Gatti:2017}.  While such priors are believed to be robust, they are difficult to obtain, require data external to the correlation function measurements, and in the case of shear bias, rely on image simulations which may not exactly match the data\footnote{As described in Section 5 of \citet{Zuntz17}, the residual shear calibration bias in \textsc{MetaCalibration} from PSF modeling errors is determined using image simulations, even though the \textsc{MetaCalibration} algorithm itself does not require simulations.}.  Because of these challenges and associated uncertainties, it would be advantageous if the correlation function measurements themselves could break the nuisance parameter degeneracies, and {\it self-calibrate} $m$ and $\Delta z_{\rm s}$.   

As pointed out by several authors \citep[e.g.][]{Vallinotto:2012,Baxter:2016,Liu:2016,Schaan:2017} joint measurements of galaxy lensing and CMB lensing correlations can enable self-calibration of both multiplicative shear bias and photometric redshift biases.  This is possible because CMB lensing and galaxy lensing are correlated, while CMB lensing is not sensitive to these two sources of systematic error, thus breaking the three-parameter degeneracy between shear bias, galaxy bias, and $A_{\rm s}$ described above.\footnote{In principle, CMB lensing could also have some form of multiplicative bias.  However, for current measurements, any multiplicative bias is expected to be much smaller than the associated statistical errorbars.}  This degeneracy breaking is illustrated with the red contours in Fig.~\ref{fig:bm_degeneracy}. In fact, either one of $\galk$ or $\sheartk$ is sufficient to break this degeneracy.  The $\sheartk$ correlation breaks this degeneracy because this quantity depends on $m$, but not on galaxy bias; it is broken by $\galk$ because this quantity depends on galaxy bias, but not on $m$. 
 
We now investigate the potential of the \5x2pt analysis to self-calibrate the shear and photo-$z$ bias parameters by replacing the fiducial priors on these parameters (in Table~\ref{tab:params}) with non-informative, flat priors.  For $m$, we use $m \in [-1, 1]$; for $\Delta z_{\rm s}$, we use $\Delta z_{\rm s} \in [-1,1]$.

The posteriors on the shear calibration parameters resulting from the \5x2pt and \3x2pt analyses for wide priors on $m$ are summarized in the left panel of Fig.~\ref{fig:mzsconstraint}.  The blue bands in that figure illustrate the level at which the \3x2pt combination is able to self-calibrate the multiplicative shear bias, roughly $\sigma(m) \sim 0.2$.  Note that the confidence intervals shown in Fig. ~\ref{fig:mzsconstraint} are not centered on the input shear values, even though the maximum likelihood point in the full parameter space does occur at the input parameter values; this is simply the result of projecting the higher dimensional parameter space to the 1D parameter space shown in the figure.  We find that the \5x2pt combination is able to significantly improve on the self-calibration of $m$, reaching constraints of roughly $\sigma (m) \sim 0.1$, with the constraints improved somewhat for higher redshift bins (orange bands).  This level of shear calibration is certainly interesting, but is not yet competitive with priors on the $m$ obtained in the fiducial \citet{DESy1:2017} \3x2pt analysis (black bands). 

Changing the priors on $\Delta z_{\rm s}$ to be flat reveals that the \5x2pt analysis constrains these biases at roughly $\sigma(\Delta z_{\rm s}) \sim 0.03-0.04$ (right panel of Fig.~\ref{fig:mzsconstraint}).  This level of constraint is only a factor of $\sim 2$ weaker than the fiducial priors on $\Delta z_{\rm s}$. However, we find that the posterior on $\Delta z_{\rm s}$ from the \3x2pt analysis is almost identical to that from \5x2pt.  The reason for this is that $\Delta z_{\rm s}$ is not impacted by the three parameter degeneracy that affects $m$ in the \3x2pt analysis, and can therefore be tightly constrained using \3x2pt alone.  

\begin{figure}
\includegraphics[width=\linewidth]{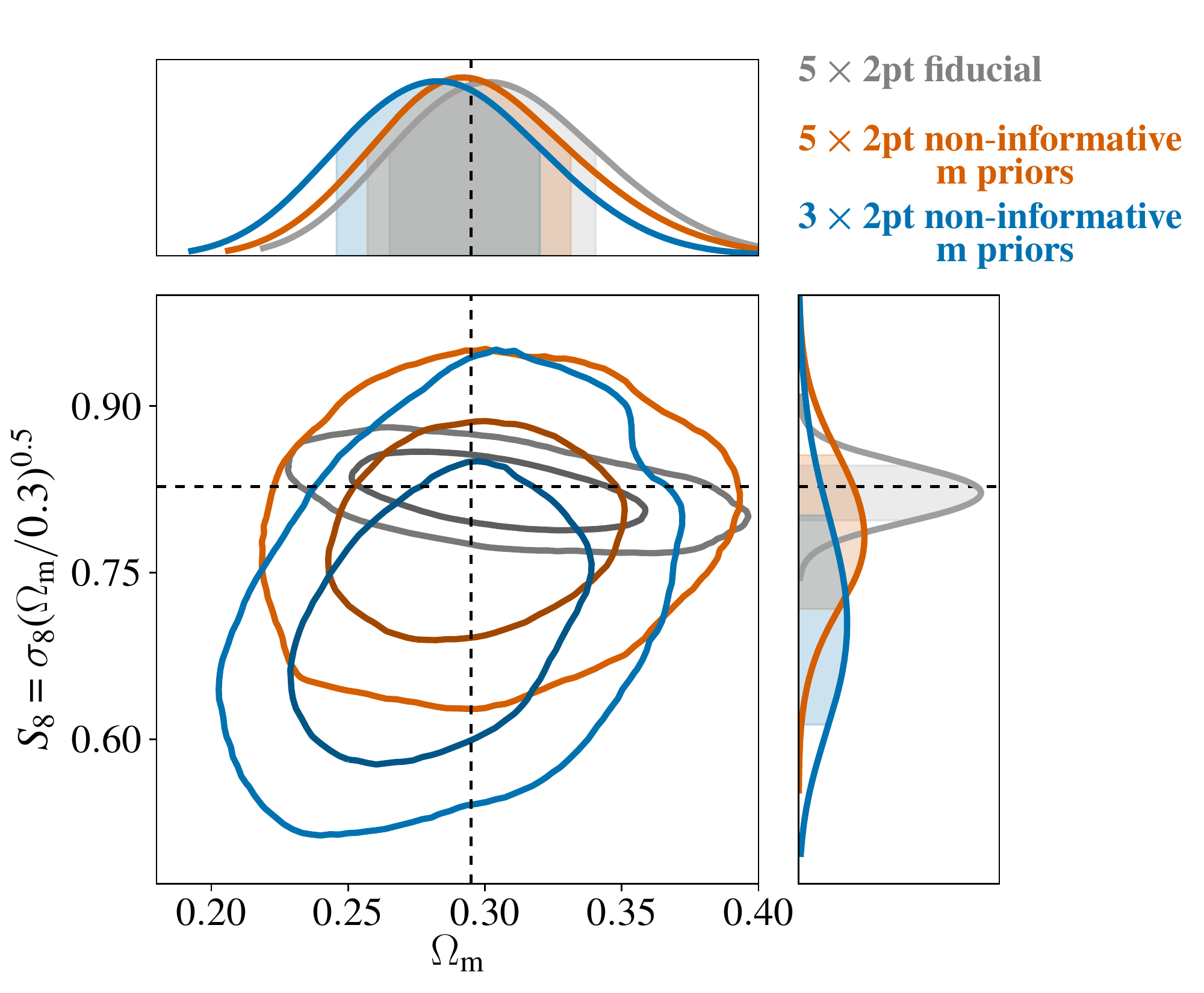}
\caption{Constraints on $\om$, $S_8$ with and without fiducial priors on shear calibration bias.  With non-informative priors on shear calibration bias, the \5x2pt analysis is able to obtain tight cosmological constraints.  The \3x2pt analysis, however, is significantly degraded in the absence of tight priors on shear calibration. }
\label{fig:cosmo_freem}
\end{figure}

\begin{figure}
\includegraphics[width=\linewidth]{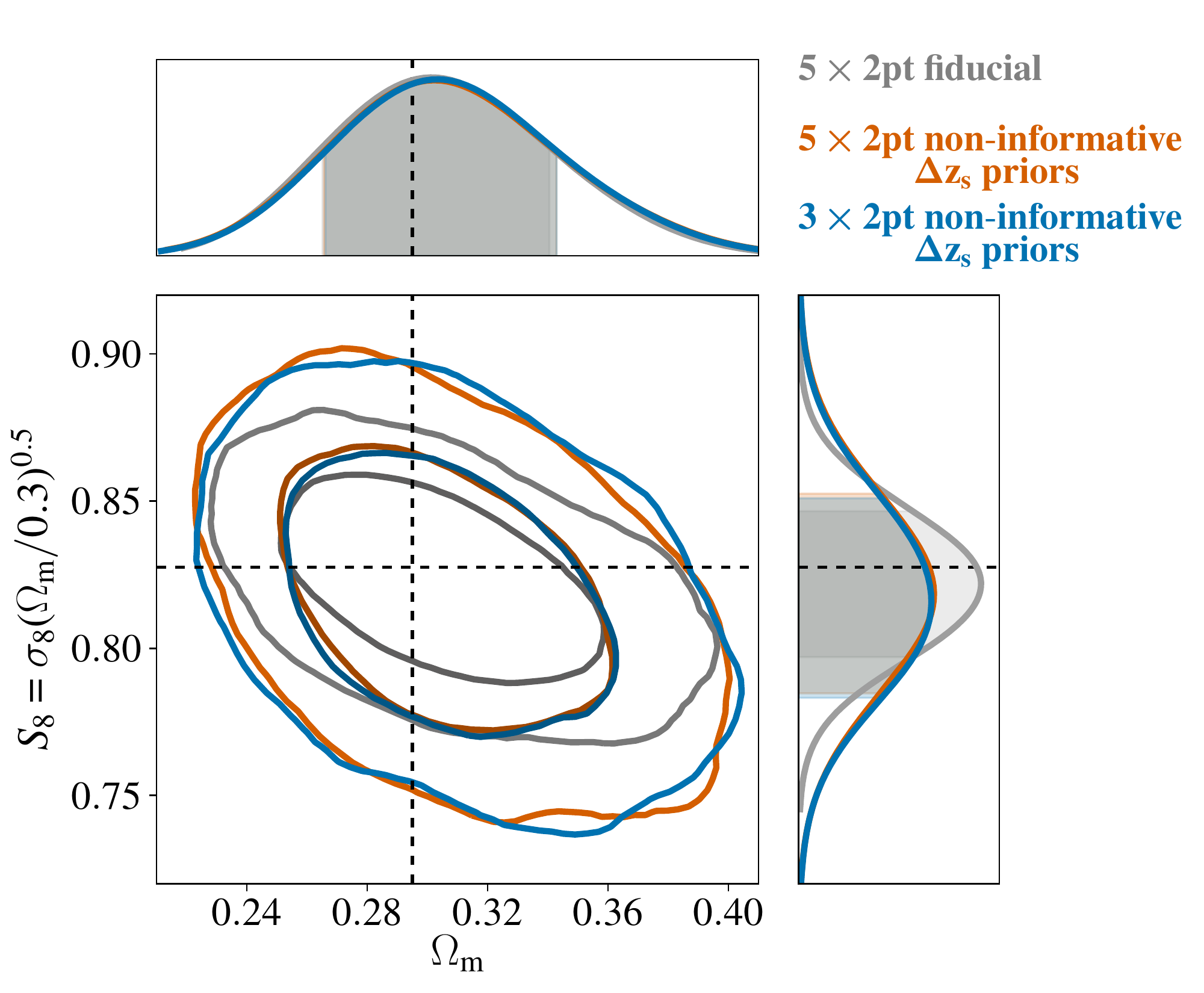}
\caption{Constraints on $\om$, $S_8$ with and without the fiducial informative priors on source photometric redshift bias.}
\label{fig:cosmo_freezs}
\end{figure}

The constraints on $\Omega_{\rm m}$ and $S_8$ obtained from the \3x2pt and \5x2pt analyses when the priors on $m$ are very wide and flat are shown in Fig.~\ref{fig:cosmo_freem}.  This figure highlights the exciting potential of the \5x2pt analysis: with a non-informative prior on $m$, the \5x2pt analysis can obtain significantly tighter cosmological constraints than the \3x2pt analysis.  We see that weakening the priors on $m$ mostly degrades the cosmological constraints in the $S_8$ direction.  This is because $S_8$ effectively controls the amplitude of the correlation functions, and it is thus strongly impacted by the degeneracy between shear calibration, galaxy bias, and $A_{\rm s}$ described above.  Also shown in Fig.~\ref{fig:cosmo_freem} are the contours obtained from the \5x2pt analysis with the fiducial $m$ priors.  Comparing these contours to those with the loose $m$ priors reveals that the priors on $m$ do contribute some information to the cosmological constraints.  This is not surprising, given that the level at which \5x2pt self-calibrates $m$ is significantly looser than the fiducial priors on $m$.  
Fig.~\ref{fig:cosmo_freezs} shows the cosmological constraints obtained from the \5x2pt and \3x2pt analyses when the priors on $\Delta z_{\rm s}$ become non-informative. In this case, we see little improvement of the \5x2pt combination relative to the \3x2pt combination.  We find that the fiducial priors on $\Delta z_{\rm s}$ are useful for improving cosmological constraints in the \5x2pt analysis, indicating that the data is not self-calibrating for this parameter.

\section{Discussion}
\label{sec:conclusion}

We have presented the methodology for jointly analyzing the combination of five two-point functions that can be formed from the combination of the $\delta_{\rm g}$, $\gamma$ and $\kappa_{\rm CMB}$ observables (not including the $\kappa_{\rm CMB}$ autocorrelation).  This methodology will be applied to a forthcoming analysis using data from DES, SPT and \Planckc.  

Essential to this analysis is the characterization of the bias in maps of $\kappa_{\rm CMB}$ induced by the thermal Sunyaev-Zel'dovich effect.  Our estimate of this bias suggests that it could be quite large at small scales.  Given the uncertainties associated with this estimate, we do not attempt to model tSZ bias in our analysis.  Instead, we remove angular scales that are estimated to be strongly affected by the bias, at the cost of increasing our statistical errorbars.  This degradation is significant: the total expected signal-to-noise of the $\sheartk$ and $\galk$ cross-correlations is roughly 20; after the scale cuts, this is reduced to 8.8.

Given the scale cuts required to remove tSZ contamination of the $\kappa_{\rm CMB}$ maps, we find that the joint cosmological constraining power of $\galk$ and $\sheartk$ is significantly weaker than the \3x2pt analysis (Fig.~\ref{fig:fid_constraints}).  Consequently, the \5x2pt analysis does not lead to dramatic improvement in cosmological constraints given the fiducial priors of the \3x2pt analysis.  

However, we find that the \5x2pt analysis can significantly improve on the cosmological constraining power of the \3x2pt analysis in the case that priors on the multiplicative shear biases are loosened.  As shown in Fig.~\ref{fig:cosmo_freem}, with essentially no information on the multiplicative bias parameters, the \5x2pt analysis can still obtain tight cosmological constraints.

Given the large degradation in signal-to-noise that results from cutting scales affected by tSZ contamination, future work to model or remove such contamination is strongly motivated.  More accurate estimates of the contamination could be achieved with high signal-to-noise and high resolution Compton $y$ maps.  Alternatively, such contamination could be removed from the $\kappa_{\rm CMB}$ maps using either multi-frequency component separation methods to remove tSZ from the CMB temperature maps, or by constructing the  $\kappa_{\rm CMB}$ maps instead from maps of the CMB polarization, since the tSZ signal is nearly unpolarized.  

\section*{Acknowledgements}

This paper has gone through internal review by the DES collaboration.

The contour plots in this paper were made using \texttt{ChainConsumer}\footnote{\texttt{https://github.com/Samreay/ChainConsumer}} \citep{Hinton2016}.

EB is partially supported by the US Department of Energy grant
DE-SC0007901.  The Melbourne group acknowledges support from the Australian Research Council's Future Fellowships scheme (FT150100074). 
PF is funded by MINECO, projects ESP2013-48274-C3-1-P,
ESP2014-58384-C3-1-P, and ESP2015-66861-C3-1-R.  ER is supported by
DOE grant DE-SC0015975 and by the Sloan Foundation, grant FG-
2016-6443.

Support for DG was provided by NASA through Einstein Postdoctoral
Fellowship grant number PF5-160138 awarded by the Chandra X-ray
Center, which is operated by the Smithsonian Astrophysical Observatory for NASA under contract NAS8-03060.

Funding for the DES Projects has been provided by the U.S. Department
of Energy, the U.S. National Science Foundation, the Ministry of
Science and Education of Spain, the Science and Technology Facilities
Council of the United Kingdom, the Higher Education Funding Council
for England, the National Center for Supercomputing Applications at
the University of Illinois at Urbana-Champaign, the Kavli Institute of
Cosmological Physics at the University of Chicago, the Center for
Cosmology and Astro-Particle Physics at the Ohio State University, the
Mitchell Institute for Fundamental Physics and Astronomy at Texas A\&M
University, Financiadora de Estudos e Projetos, Funda{\c c}{\~a}o
Carlos Chagas Filho de Amparo {\`a} Pesquisa do Estado do Rio de
Janeiro, Conselho Nacional de Desenvolvimento Cient{\'i}fico e
Tecnol{\'o}gico and the Minist{\'e}rio da Ci{\^e}ncia, Tecnologia e
Inova{\c c}{\~a}o, the Deutsche Forschungsgemeinschaft and the
Collaborating Institutions in the Dark Energy Survey.

The Collaborating Institutions are Argonne National Laboratory, the
University of California at Santa Cruz, the University of Cambridge,
Centro de Investigaciones Energ{\'e}ticas, Medioambientales y
Tecnol{\'o}gicas-Madrid, the University of Chicago, University College
London, the DES-Brazil Consortium, the University of Edinburgh, the
Eidgen{\"o}ssische Technische Hochschule (ETH) Z{\"u}rich, Fermi
National Accelerator Laboratory, the University of Illinois at
Urbana-Champaign, the Institut de Ci{\`e}ncies de l'Espai (IEEC/CSIC),
the Institut de F{\'i}sica d'Altes Energies, Lawrence Berkeley
National Laboratory, the Ludwig-Maximilians Universit{\"a}t
M{\"u}nchen and the associated Excellence Cluster Universe, the
University of Michigan, the National Optical Astronomy Observatory,
the University of Nottingham, The Ohio State University, the
University of Pennsylvania, the University of Portsmouth, SLAC
National Accelerator Laboratory, Stanford University, the University
of Sussex, Texas A\&M University, and the OzDES Membership Consortium.

Based in part on observations at Cerro Tololo Inter-American
Observatory, National Optical Astronomy Observatory, which is operated
by the Association of Universities for Research in Astronomy (AURA)
under a cooperative agreement with the National Science Foundation.

The DES data management system is supported by the National Science
Foundation under Grant Numbers AST-1138766 and AST-1536171.  The DES
participants from Spanish institutions are partially supported by
MINECO under grants AYA2015-71825, ESP2015-88861, FPA2015-68048,
SEV-2012-0234, SEV-2016-0597, and MDM-2015-0509, some of which include
ERDF funds from the European Union. IFAE is partially funded by the
CERCA program of the Generalitat de Catalunya.  Research leading to
these results has received funding from the European Research Council
under the European Union's Seventh Framework Program (FP7/2007-2013)
including ERC grant agreements 240672, 291329, and 306478.  We
acknowledge support from the Australian Research Council Centre of
Excellence for All-sky Astrophysics (CAASTRO), through project number
CE110001020.

This manuscript has been authored by Fermi Research Alliance, LLC
under Contract No. DE-AC02-07CH11359 with the U.S. Department of
Energy, Office of Science, Office of High Energy Physics. The United
States Government retains and the publisher, by accepting the article
for publication, acknowledges that the United States Government
retains a non-exclusive, paid-up, irrevocable, world-wide license to
publish or reproduce the published form of this manuscript, or allow
others to do so, for United States Government purposes.

The South Pole Telescope program is supported by the
National Science Foundation through grant PLR-1248097.
Partial support is also provided by the NSF Physics Frontier
Center grant PHY-0114422 to the Kavli Institute of Cosmological
Physics at the University of Chicago, the Kavli
Foundation, and the Gordon and Betty Moore Foundation
through Grant GBMF\#947 to the University of Chicago.
Argonne National Laboratory’s work was supported
under the U.S. Department of Energy contract DE-AC02-
06CH11357.

Argonne National Laboratory work was supported under U.S. Department of Energy contract DE-AC02-06CH11357.

\appendix

\bibliography{thebibliography.bib}

\label{lastpage}

\end{document}